\documentclass[11pt]{article}
\usepackage{natbib}
\usepackage[english]{babel}
\usepackage[latin1]{inputenc}
\usepackage{lmodern}
\usepackage[T1]{fontenc}
\usepackage{amssymb,amsmath,amsthm,latexsym,amsfonts,amscd,dsfont,enumerate,bbm}
\usepackage{a4wide,graphicx,tikz}

\usetikzlibrary{arrows,shapes,shadows,fit,backgrounds}
\tikzstyle{trader} = [circle, draw, top color=white, bottom color=blue!30, draw=blue!50!black!100, drop shadow, minimum height=4em]
\tikzstyle{bank} = [rectangle, draw, top color=white, bottom color=red!20, draw=red!50!black!100, drop shadow, rounded corners, minimum height=3em, text width=4em, text centered]
\tikzstyle{market} = [rectangle, draw, top color=white, bottom color=green!20, draw=green!50!black!100, drop shadow, rounded corners, minimum height=3em, text width=4em, text centered]
\tikzstyle{background} = [rectangle,fill=gray!10, inner sep=0.2cm, rounded corners=5mm]
\tikzstyle{line} = [draw, latex'-latex']
\tikzstyle{from} = [draw, latex'-]
\tikzstyle{to} = [draw, -latex']

\usepackage[]{hyperref}
 \hypersetup{
 colorlinks=true, 
 breaklinks=true, 
 urlcolor= blue, 
 linkcolor= blue, 
 bookmarksopen=true, 
 pdfauthor={Pallavicini, A.}
 }

\newtheorem{theorem}{Theorem}[section]
\newtheorem{remark}[theorem]{Remark}

\newcommand{\Eq}[1]{{\[{#1}\]}}

\newcommand{\Ex}[2]{\mathbb{E}_{#1}\!\left[\,#2\,\right]}

\newcommand{\ExT}[3]{\mathbb{E}_{#1}^{#2}\!\left[\,#3\,\right]}

\newcommand{\ind}[1]{1_{\{#1\}}}

\newcommand{\cov}[4]{{\rm Cov}_{#1}^{#2}\!\left[\,#3\,,\,#4\,\right]}

\usepackage{eurosym}

\title{FX Modelling in Collateralized Markets:\\ foreign measures, basis curves, and pricing formulae }
\author{
Nicola Moreni\thanks{Banca IMI Milan, {\tt nicola.moreni@bancaimi.com}}
\ \ \
Andrea Pallavicini\thanks{Imperial College London and Banca IMI Milan, {\tt a.pallavicini@imperial.ac.uk}}
}
\date{
\small First Version: October 28, 2014.  This version: \today
}

\begin{document}

\maketitle

\begin{abstract}
We present a general derivation of the arbitrage-free pricing framework for multiple-currency collateralized products. We include the impact on option pricing of the policy adopted to fund in foreing currency, so that we are able to price contracts with cash flows and/or collateral accounts expressed in foreign currencies inclusive of funding costs originating from dislocations in the FX market. Then, we apply these results to price cross-currency swaps under different market situations, to understand how to implement a feasible curve bootstrap procedure. We present the main practical problems arising from the way the market is quoting liquid instruments: uncertainties about collateral currencies and renotioning features. We discuss the theoretical requirements to implement curve bootstrapping and the approximations usually taken to practically implement the procedure. We also provide numerical examples based on real market data.   
\end{abstract}

{\bf JEL classification code: G13. \\ \indent AMS classification codes: 60J75, 91B70}

\medskip

{\bf Keywords:} Arbitrage-Free Pricing, Collateral, Collateral Convexity, Funding Costs, Funding Policy, Foreign Currency, FX Market, FX Swap, Cross-Currency Swap, Curve Bootstrapping, Multiple Currencies, Currency Triplets.

\newpage
{\small \tableofcontents}
\vfill
{\footnotesize \noindent The opinions here expressed  are solely those of the authors and do not represent in any way those of their employers.}
\newpage

\pagestyle{myheadings} \markboth{}{{\footnotesize  N. Moreni, A. Pallavicini, FX Modelling in Collateralized Markets}}

\section{Introduction}
\label{sec:introduction}

Since the financial crisis of 2007 banks and financial institutions, which were so far considered as non-defaultable corporations, started being suspicious about the liquidity availability and credit worthiness of their counterparties. Borrowing money, even for short maturities (under one year), became more expensive, as banks charged their counterparties higher rates for unsecured lending. The shortage of funding sources forced central banks to adopt a number of non-standard measures to support financing conditions and credit flows both in domestic and foreign currencies.

Despite these efforts, market frictions and dislocations, which were already present before the crisis, strenghtened. This happened both in single-currency money markets and in FX swap markets. In particular, cross-currency absence-of-arbitrage relationships involving market quotes of FX forward rates and single-currency zero-coupon bonds displayed severer violations. This problem is discussed in \cite{Baba2008}, where the authors search for an explanation of the failure of the covered interest parity conditions between USD and EUR, GBP and JPY during the crisis period. They identify three causes: (i) the market perceiving European financial institutions more risky than US ones, (ii) the shortage in US dollars of non-US financial institutions leading to one-sided order flows concentrated on US dollar borrowing, and (iii) the difficulty to size the borrowing costs in the money market by means of the Libor rate. Their analysis is completed in \cite{Baba2009} with the discussion of the effects of central bank's policies to contrast the liquidity shortage. The evidence of a positive premium paid by non-US financial institutions to fund in US dollars is also discussed in \cite{Coffey2009}, \cite{Ossolinski2010}, \cite{Mancini2011}, and \cite{Filipozzi2013}. Moreover, the presence of different time zones around the world also contributes to the segmentation across currencies. The Payments Risk Committee (\cite{PrcFed2012}), in a research sponsored by the Federal Reserve Bank of New York, tracked the USD intraday flows among financial institutions and clearing banks. The committee found relative peaks at the beginning and end of primary eastern U.S. business hours. As clearing houses and FX settlement institutions impose to settle payments at specified times, these USD liquidity peaks do not correspond to the time frame during which European financial institutions are obliged to fulfill USD payments. Hence, European players experience a relative shortage of USD.

The failure of covered interest parity has a direct consequence in derivative option prices. An investor, funding derivative contracts and hedging instruments along with their collateral accounts, requires liquidity in one or more currencies. Cash in foreign currencies is usually obtained by trading FX spot and swap contracts. Thus, market dislocations may produce additional costs in funding and hedging activities and, during turbulent periods, can also lead to severe liquidity shortages as shown in \cite{Barkbu2010}.

We notice that these funding costs depend on the particular funding strategy adopted by the investor. Indeed, there are different ways to raise money in a foreign currency. For instance, a domestic institution could issue debt notes denominated in a currency other than its domestic one, entering into a loan whose interest and capital repayment at maturity will equally be expressed as amounts of that currency. The actual funding policy adopted by an institution is a collection of different strategies, driven not only by financial factors. Thus, to introduce an arbitrage-free pricing framework we need to select a particular funding policy. This is a problem which is usually faced in funding cost literature, as in \cite{Perini2011} or \cite{Crepey2011}, and addressed also in practitioner conferences, as in \cite{Kyaer2015}. Here, we explicitly assume that a domestic investor can fund in foreign currencies only by means of FX swaps. Thus, prices of derivative contracts with cash flows or collateral accounts expressed in foreign currencies should include funding costs originating from the FX swap market, as shown in \cite{Piterbarg2012}. However, brokers currently quote FX and cross-currency swaps (CCS) regardless of the collateralization policy, which may as well be absent.

In general, the FX market does not quote instruments sufficient to fix all the degrees of freedom of dynamical models describing the relevant financial risks. For instance, FX swaps are not actively traded for mid-long maturities, while the market standard for CCS has a non-linear payoff in interest and FX rates.  Moreover, the market of cross-currency products is essentially USD based, so that we need to perform triangulations to connect currencies for which no quotes are available. For these reasons, in order to price exotic products, market participants are forced to make rough assumptions whose adequacy should be better understood and sized. This paper sets within this context and aims to shed some light both from a theoretical and a market practice point of view. 

First, we describe how to coherently price derivatives with flows and/or collateral posting in different currencies in presence of market dislocations and relying on funding strategies based on FX swaps. We extend the usual arbitrage-free pricing framework to accommodate collateral accounts by means of a more general definition of dividend and gain processes and we give clear definitions of the relevant pricing measures. In doing so we adapt the results of \cite{Perini2011} and \cite{Crepey2011} to multiple currencies in case of perfect collateralization. The previous works of \cite{Piterbarg2012}, \cite{McCloud2013} and \cite{Gimenez2014} do not attempt to formulate a generic pricing framework, while the papers by \cite{Fujii2013,Fujii2015} derive a pricing framework which does not include the impact of market dislocations. We apply these results to derive pricing formulae for derivative contracts under different collateralization agreements. 

A second contribution of the present work is to assess the validity of the approximations usually done in pricing FX instruments because of the lack of a sufficiently rich set of market quotes. Market practice will be analyzed both in terms of the theory we formulated and by means of numerical examples. In particular, we will discuss CCS pricing methodologies.

\medskip

The structure of the paper is the following. In Section~\ref{sec:pricing} we present the pricing framework and we derive  generic pricing formulae for different combinations of cash flow and collateral currencies. The formal derivation of the pricing framework is given in the appendix. In Section~\ref{sec:bootstrap} we apply the results to the pricing of FX swaps and CCS, and we discuss curve bootstrapping. Some calculations are derived in the appendix. In Section~\ref{sec:effectiveDisc} we investigate some approximations usually done in the practice when evaluating CCS and discuss their validity by means of numerical examples. In Section~\ref{sec:conclusion} we review our contributions and hint for further developments.

\section{Pricing with Different Currencies under Collateralization}
\label{sec:pricing}

After the crisis of 2007 the default events of the counterparties cannot be ignored any longer. Furthermore, liquidity basis are now present in all markets to reflect the difficulties in raising cash and assets. A direct consequence is the diffusion of collateralization procedures to remove counterparty credit risk, either by means of bilateral agreements, as the ISDA CSA, or by trading with centralized counterparties (CCP).

\subsection{The Impact of Collateralization in Pricing}

Collateralization means the right of recourse to some asset of value that can be sold or the value of which can be applied as a guarantee in the event of default on the transaction. In case of no default happening, at maturity the collateral provider expects to get back the remaining collateral from the collateral taker. Similarly, in case of default happening earlier, after netting the collateral account with the cash flows of the transaction, the collateral provider expects to get back the remaining collateral on the account, if any.

A margining or collateralization procedure consists in a pre-fixed set of dates during the life of a deal when both parties post or withdraw collateral amounts, according to their current exposure, to or from an account held by the collateral taker. We name collateral taker the party which is receiving the collateral assets, the other one being the collateral giver. Moreover, the procedure ensures that the collateral taker remunerates the account at a particular accrual rate.

The whole mechanism is designed so to return to the owner all the assets posted as collaterals when the collateralized trade ends. Yet, the interests paid by the collateral taker are not reimbursed, so that they must be included in the pricing framework used to evaluate the contract. We can think of such interests as a stream of dividends to be added to the contractual cash flows of the deal, as described in \cite{BrigoCapponiPallaviciniPapatheodorouIJTAF}.

To proceed in pricing FX derivatives we need an extension of the pricing framework to include collateralization. We develop it in appendix \ref{sec:appendix}. Here, we focus on the main result, namely the pricing formula for perfectly collateralized contracts.

We call a contract perfectly collateralized when we can assume that the collateralization procedure is able to prevent any loss in case of default of one of the two counterparties. A good approximation of perfect collateralization is a contract with collateral assets exchanged each day with a liquid market allowing to unroll the deal in case of default as quick as possible without further losses. Usually this approximation is used for interest-rate derivatives and most of FX derivatives. We discuss these problems in the appendix.

Here, we consider a contract perfectly collateralized which is hedged with perfectly collateralized instruments. The collateral account of the deal is remunerated at the accrual rate $c_t$. The price of the contract, see equation \eqref{eq:perfect}, is given by
\begin{equation}
\label{eq:pricing}
V_t = \int_t^T \Ex{t}{ D(t,u;r) \left( d\pi_u + V_u (r_u - c_u) \,du \right) } = \int_t^T \Ex{t}{ D(t,u;c) \,d\pi_u }
\end{equation}%
where the expectation is taken under the martingale measure $\mathbb Q$, which is equivalent to the physical world one and  ensures the absence of arbitrages, as detailed in the appendix. The discount factor for a generic rate $x_t$ is defined as
\[
D(t,T;x) := \exp\left\{-\int_t^T du\, x_u\right\}
\]%
and the coupon process $\pi_t$ is defined as
\[
\pi_t := \sum_{i=1}^N \gamma_i \ind{T_i \le t}
\]%
where $\gamma_i$ is the coupon amount paid at time $T_i$.

When we trade a derivative on the market we have to fund its contractual cash flows along with the margining procedure. Moreover, we have to fund also the hedging instruments we trade to remove or reduce the risks of the derivative. When all the obligations are expressed in the currency of the risk-free bank account we are using as numeraire, we can use directly equation \eqref{eq:pricing} to calculate prices.

Notice that equation \eqref{eq:pricing} does not depend on the risk-free rate\footnote{This is a general property of pricing equations inclusive of collateral and funding costs, first noticed in \cite{Perini2011}. A complete discussion can be found in \cite{brigo2015nonlinear}.}. When a derivative contract is perfectly collateralized, we can price it as it was funded by means of the collateral account. Thus, if we consider only perfectly collateralized deals in a single currency, we can identify the funding curve with the curve of collateral accrual rates.

For instance, if we consider the Euro money market, we have that most of the contracts are collateralized on a daily basis with accrual rate equal to the overnight rate $e_t$, so that, by applying the perfect collateralization approximation, we can write
\[
V_t = \int_t^T \Ex{t}{ D(t,u;e) \,d\pi_u }\,.
\]%

We can cast the above equation into a more explicit form if we define the overnight domestic funding curve as
\begin{equation}
\label{eq:zcbond}
P_t(T;e) := \Ex{t}{D(t,T;e)}
\end{equation}%
Then, we define the collateralized domestic $T$-forward measure ${\mathbb Q}^{T;e}$ by means of the following Radon-Nikodym derivative.
\begin{equation}
\label{eq:forwardrnd}
Z_t(T;e) := \left.\frac{d \mathbb{Q}^{T;e}}{d \mathbb{Q}}\right|_t := \frac{\Ex{t}{D(0,T;e)}}{P_0(T;e)} = \frac{D(0,t;e) P_t(T;e)}{P_0(T;e)}
\end{equation}%
which is a ${\mathbb Q}$ martingale and it is normalized so that $Z_0(T;e)=1$. If we switch to the collateralized $T$-forward measure ${\mathbb Q}^{T;e}$, we obtain
\[
V_t = \int_t^T P_t(u;e) \,\ExT{t}{u}{ \,d\pi_u }\,.
\]%

\subsection{Funding Strategies in Domestic and Foreign Currencies}

If some cash flow is expressed in a different currency, we should describe how the investor can obtain cash in such currencies to fulfill the contractual agreements. In general, we could refer to such problem as the problem of funding in different currencies. In the following it is crucial to assume that the investor can fund without relevant restrictions in one particular currency by accessing a risk-free bank account, and we call such currency the domestic currency. All the other currencies are called foreign currencies.

The domestic currency can be also thought to as the one in which the investor's balance sheet is written. As discussed in the introduction, the presence of asynchronous trading windows and the limited access to on-shore liquidity channels for off-shore institutions, create market segmentation between currencies. Hence, the problem we describe here may naturally lead to asymmetrical evaluation of financial contracts. Funding in foreign currencies requires to trade market instruments paying cash flows in such currencies, and, if required, to remunerate their collateral accounts in the proper currency. We insist on the description of the market strategy used to implement funding in foreign currencies, since the collateralization procedures required by the strategy will affect the pricing formulae.

\subsubsection{Accessing the Foreign Money Market}

In order to grant access to foreign currencies through inter-dealer trades, the FX money market quotes two instruments which are commonly used to implement funding strategies in foreign currencies: the FX spot and forward (or outright) contracts. Morevoer, the market quotes also combinations of a long (short) FX spot contract and of a short (long) FX forward contract, usually named FX swap.

The FX spot contract allows to exchange a lump of money expressed in one currency into an equivalent value denominated in the other one. The ratio of the two quantities is the FX rate. The FX spot contract usually settles the operation in two business days. If a longer maturity is required, the investor can trade FX forwards, which allow to lock in a specific FX rate level at a future date, alternatively we can use a combination of FX spot and swap contracts. In the following we focus on FX swap contracts.

A FX swap is a contract in which the investor borrows cash from the counterparty in foreign currency while lending domestic currency to the same party. At inception, namely at spot date, one unit of domestic currency is exchanged against the equivalent amount of foreign currency, while at maturity one unit of domestic currency is exchanged back against a given quantity of foreign currency that was determined at inception by market bid-ask dynamics. FX swap contracts are margined daily. We assume that the investor is able to find on the market FX swaps requiring a collateral account remunerated in domestic currency at a contractual accrual rate. In the Euro area the collateral rate is usually set equal to the EONIA rate, namely the reference rate for the inter-bank overnight unsecured deposits.

Let $\chi_t$ be the FX market rate converting one unit of foreign currency into a quantity of domestic currency as seen at time $t$ and let $e_t$ be the domestic collateral accrual rate for FX products. A FX swap contract started at $t$ and collateralized at $e$ will exchange, at its maturity $T$, one unit of domestic currency against $1/X_t(T;e)$ units of foreign currency. The quantity $X_t(T;e)$ correspond to the market quote for the given FX swap\footnote{Brokers indeed do quote the difference $\delta_t(T;e):=X_t(T;e)-\chi_t.$}.

Then, we use FX swap contracts to build funding strategies in a foreign currency. In particular, if we assume that the margining procedure occurs on a continuous time basis and it is able to remove all credit risk (perfect collateralization), we can use the pricing formula \eqref{eq:pricing} to get
\begin{equation}
\label{eq:fxswap}
V^{\rm FXswap}_t := \Ex{t}{ \left(\frac{\chi_T}{X_t(T;e)} - 1\right) D(t,T;e) }\,,
\end{equation}%
where the price of the contract does not depend on the initial cash flows at time $t$ since they offset. The FX forward rate is determined to sell the FX swap contract at par, so that we can solve the above equation w.r.t. the FX forward rate to get
\begin{equation}
\label{eq:fxforward}
X_t(T;e) = \frac{ \Ex{t}{\chi_T D(t,T;e)} }{ \Ex{t}{D(t,T;e)} } = \ExT{t}{T;e}{\chi_T}\,.
\end{equation}%
Notice that FX forward rates depend on collateral rates, and, as a consequence, FX forward rates observed in instruments with different collateralization are different.

\begin{remark}{\bf (FX swap contracts with foreign reference leg)}
The reference leg of a FX swap contract can be expressed in foreign currency, namely we can consider a FX swap contract where at inception one unit of foreign currency is exchanged against the equivalent amount of domestic currency, while at maturity one unit of foreign currency is exchanged back against a given quantity of domestic currency. If we still assume domestic collateralization at overnight rate $e_t$, we can apply equation \eqref{eq:pricing} to obtain
\[
{\tilde V}^{\rm FXswap}_t := \Ex{t}{ \left(\chi_T - {\tilde X}_t(T;e)\right) D(t,T;e) }
\]%
where ${\tilde X}_t(T;e)$ is the par rate of the contract, so that
\[
{\tilde X}_t(T;e) = \frac{ \Ex{t}{ \chi_T D(t,T;e)} }{ \Ex{t}{ D(t,T;e) } } = \ExT{t}{T;e}{ \chi_T } = X_t(T;e)\,.
\]%
Thus, we get that this forward rate is exactly the same as the one given in equation \eqref{eq:fxforward}.
\end{remark}

\subsubsection{Collateralized Foreign Measure}

We can use FX forward rate to define a new pricing measure. We define the collateralized foreign measure ${\mathbb Q}^{b}$ by means of the following Radon-Nikodym derivative
\begin{equation}
\label{eq:basisrnd}
Z^f_t(e) := \left.\frac{d \mathbb{Q}^{b}}{d \mathbb{Q}}\right|_t := \frac{\chi_t}{\chi_0} D(0,t;e-b^f(e))
\end{equation}%
where we define the basis rate $b^f_t(e)$ 
\begin{equation}
\label{eq:basisrate}
b^f_t(e) \,dt := e_t \,dt - \Ex{t}{\frac{d\chi_t}{\chi_t}}\,.
\end{equation}%
Notice that $Z^f_t(e)$ is a ${\mathbb Q}$ martingale normalized so that $Z^f_0(e)=1$.

Hence, if we use the above measure in the definition of FX forward rate, we get
\begin{equation}\label{eq:fromQtoQb}
\ExT{t}{b}{D(t,T;b^f(e))} = \frac{1}{\chi_t }\Ex{t}{\chi_T D(t,T;e)}\,.
\end{equation}
The above equation provides us a very clear interpretation of the cost of funding a cash flow of one unit of foreign currency via domestic collateralization. Actually, it states that the result, expressed in foreign currency units, is the same as discounting the flow by means of the basis collateral rate $b^f(e)$ under an appropriate measure ${\mathbb Q}^{b}$.

We can write the above equation in more explicit form if we define the effective foreign funding curve as
\begin{equation}
\label{eq:basiszcbond}
P^f_t(T;e) := \ExT{t}{b}{D(t,T;b^f(e))}\,.
\end{equation}%
Thus, we obtain
\begin{equation}
\label{eq:switch}
P^f_t(T;e) = \frac{X_t(T;e)}{\chi_t}P_t(T;e)\,.
\end{equation}%
We stress again that the funding discounting curves depends on the accrual rate of the collateral accounts required by the funding instruments.

For later user we can define also the collateralized foreign $T$-forward measure ${\mathbb Q}^{T;b}$ by means of the following Radon-Nikodym derivative.
\begin{equation}
\label{eq:basisforwardrnd}
Z^f_t(T;e) := \left.\frac{d \mathbb{Q}^{T;b}}{d \mathbb{Q}^{b}}\right|_t := \frac{\ExT{t}{b}{D(0,T;b^f(e))}}{P^f_0(T;e)} = \frac{D(0,t;b^f(e)) P^f_t(T;e)}{P^f_0(T;e)}
\end{equation}%
which is a ${\mathbb Q}^{b}$ martingale and it is normalized so that $Z^f_0(T;e)=1$.

We continue with the derivation of pricing formulae for various combinations of domestic and foreign cash flows and collateral assets.

\subsection{Derivation of Pricing Formulae}

When dealing with foreign currencies we have to face different pricing problems according to the currency of contractual cash flows and collateral accounts. Here we review the results, referring to the appendix for a formal derivation. 

We consider three cases: (i) domestic coupon contracts with collateral posted in a foreign currency, (ii) contracts with cash flows denominated in a foreign currency but domestic collateral, (iii) contracts with foreign cash flows and collateral. In Table~\ref{tab:pricingformula} we summarize the results.

\begin{table}[t]
\begin{center}
\renewcommand\arraystretch{2}
\begin{tabular}{|c|c||l|}\hline
$\pi_t$ & $C_t$ & Pricing Formula \\\hline\hline
d & d & $V_t = \int_t^T \Ex{t}{ D(t,u;c) \,d\pi_u }$ \\\hline
d & f & $V_t = \int_t^T \Ex{t}{ D(t,u;c^f - b^f(e) + e) \,d\pi_u }$ \\\hline
f & d & $V^f_t = \int_t^T \ExT{t}{b}{ D(t,u;c + b^f(e) - e) \,d\pi^f_u }$ \\\hline
f & f' & $V^f_t = \int_t^T \ExT{t}{b}{ D(t,u;c^{f'}-b^{f'}(e)+b^f(e)) \,d\pi^f_u }$ \\\hline
\end{tabular}
\renewcommand\arraystretch{1}
\end{center}
\caption{\label{tab:pricingformula}{\small
Pricing formulae for derivative contracts with domestic (d) or foreign (f or f') contractual coupons $\pi_t$ and/or collateral accounts $C_t$. Cash flows in the foreign currencies are always funded by means of FX swaps with domestic collateralization with accrual rate equal to the overnight rate $e_t$.}}
\end{table}

\subsubsection{Pricing Domestic Contracts Collateralized in Foreign Currency}

We consider the case of a derivative collateralized with assets in foreign currency remunerated at $c^f_t$ rate. We have to evaluate the cost of carry of the collateral account in foreign currency. The collateral taker must remunerate the collateral assets posted in foreign currency at the contractual rate $c^f_t$, while he is funding in the domestic currency with the risk-free bank account.

In appendix~\ref{sec:appendix} we derive the pricing formulae for this case in a more rigorous way. Here, we give a description of the problem by detailing the funding strategy followed by the collateral taker to remunerate the account. At each collateralization time $t$ we have to remunerate the collateral account, so that at $t+\Delta t$ we must have
\[
C^f_t (1 + c^f_t \Delta t)
\]%
in the collateral account, where $c^f_t$ is the derivative collateral rate. In order to obtain such foreign cash, we enter at time $t$ into a FX swap with notional
\[
X_t(t+\Delta t;e) C^f_t (1 + c^f_t \Delta t)\,.
\]%
On the other hand the FX swap require to pay back the notional in domestic currency at $t+\Delta t$, which we can fulfill by entering at time $t$ into a risk-free zero-coupon bond with notional
\Eq{
P_t(t+\Delta t;r) X_t(t+\Delta t;e) C^f_t (1 + c^f_t \Delta t)
}%
where $P_t(T;r):=\Ex{t}{D(t,T;r)}$ is the price of the risk-free zero-coupon bond. Thus, the dividend to be payed at each margining date is equal to
\Eq{
\chi_t C^f_t - P_t(t+\Delta t;r) X_t(t+\Delta t;e) C^f_t (1 + c^f_t \Delta t)\,.
}%
We can solve the FX forward rate in term of the basis curve by using equation \eqref{eq:switch}, and we obtain
\Eq{
\chi_t C^f_t \left( 1 - P_t(t+\Delta t;r) \frac{P^f_t(t+\Delta t;e)}{P_t(t+\Delta t;e)} (1 + c^f_t \Delta t) \right)\,.
}%
In the limit of small time intervals $\Delta t$ we get a continuous dividend equal to
\Eq{
\chi_t C^f_t ( r_t - c^f_t + b^f_t(e) - e_t ) \Delta t
}%
where $b^f_t(e)$ is given by equation \eqref{eq:basisrate}.

If we assume perfect collateralization, namely $V_t \doteq \chi_t C^f_t$, we can substitute the collateral costs in \eqref{eq:pricing} with the above expression to obtain the following proposition. The discount rate for a derivative perfectly collateralized in foreign currency with CSA accrual rate given by $c^f_t$, and funded by means of FX swaps collateralized at the overnight rate $e_t$, is given by $c^f_t - b^f_t(e) + e_t$, so that in case of perfect collateralization we get
\begin{eqnarray}
\label{eq:pricingforeign}
V_t &=& \int_t^T \Ex{t}{ D(t,u;r) \left( d\pi_u + V_u (r_u - c^f_u + b^f_u(e) - e_u) \,du \right) } \nonumber\\
    &=& \int_t^T \Ex{t}{ D(t,u;c^f-b^f(e)+e) \,d\pi_u }\,.
\end{eqnarray}%
This is the same result obtained in a more rigorous way in appendix \ref{sec:appendix} as given by equation \eqref{eq:perfectforeign}.

\subsubsection{Pricing Foreign Contracts Collateralized in Domestic Currency}

Contracts expressed in foreign currencies can be priced by using the collateralized foreign measure ${\mathbb Q}^{b}$ introduced by equation \eqref{eq:basisrnd}.

We consider a foreign derivative collateralized in domestic currency accruing at $c_t$ rate, and funded by means of FX swaps whose collateral account is remunerated at domestic overnight rate $e_t$. The contractual coupons can be converted at each payment dates at the spot FX rate, and we obtain from equation \eqref{eq:pricing}
\[
V_t = \int_t^T \Ex{t}{ D(t,u;c) \chi_u \,d\pi^f_u }\,.
\]%
We change measure to the collateralized foreign measure ${\mathbb Q}^{b}$, and we get
\begin{equation}
\label{eq:pricingcouponforeign}
V^f_t := \frac{V_t}{\chi_t} = \int_t^T \ExT{t}{b}{ D(t,u;c+b^f(e)-e) \,d\pi^f_u }
\end{equation}%
which can be simplified for contracts with collateral rate equal to the overnight rate in the following form.
\[
V^f_t \doteq \int_t^T \ExT{t}{b}{ D(t,u;b^f(e)) \,d\pi^f_u } = \int_t^T P^f(t,u;e) \,\ExT{t}{u;b}{ d\pi^f_u }
\;,\quad
c_t \doteq e_t
\]%
where last expectation on the right-hand side is computed under the basis forward measure $\mathbb{Q}^{T;b}.$

\subsubsection{Pricing Foreign Contracts Collateralized in Foreign Currency}

As last case we consider a foreign coupon derivative collateralized in another foreign currency accruing at $c^{f'}_t$ rate, and funded by means of FX swaps whose collateral account is remunerated at domestic overnight rate $e_t$. The contractual coupons can be converted at each payment dates at the spot FX rate, and we obtain from equation \eqref{eq:pricingforeign}
\[
V_t = \int_t^T \Ex{t}{ D(t,u;c^{f'}-b^{f'}(e)+e) \chi_u \,d\pi^f_u }\,.
\]%
We change measure to the collateralized foreign measure ${\mathbb Q}^{b}$, and we get
\begin{equation}
\label{eq:pricingforeigncouponforeign}
V^f_t = \int_t^T \ExT{t}{b}{ D(t,u;c^{f'}-b^{f'}(e)+b^f(e)) \,d\pi^f_u }
\end{equation}%
which can be simplified for contracts collateralized in the same foreign currency of the cash flows in the following form.
\[
V^f_t \doteq \int_t^T \ExT{t}{b}{ D(t,u;c^f) \,d\pi^f_u }
\;,\quad
f \doteq f'\,.
\]%

\medskip

\noindent We have seen how the choice of the collateral and cash-flow currencies modifies the pricing equation. In the following section we apply these results to understand the impact of a change of collateralization currency in derivative pricing. In particular, we will focus on FX swap and CCS pricing.

\section{Pricing FX market instruments}
\label{sec:bootstrap}
In the previous sections we deduced pricing formulae to be used by a domestic investor to value derivatives with flows and/or collateral in a foreign currency. The aim of present section is to apply the theory just developed to the pricing of FX market products, with a special focus on FX swaps and CCS. As for the single-currency interest rate markets, we find quotes only for a given set of instrument typologies and for standardized expiry/maturity dates, whereas traders need to price and hedge off-market products, with customized features. The most straightforward way to achieve this task is to bootstrap\footnote{We refer to \cite{Henrard2014} and the refrence within for a review of single-currency standard bootstrap procedures.}  a set of convenient discounting and forwarding curves, able to take into account collateral posting in foreign currencies. In terms of availability and liquidity, the instruments that can be used to calibrate those curves are FX swaps for short to mid maturities and CCS with notional resetting for mid to long maturities.  
Within this section, we first assess the impact of collateral choice on FX swaps, in order to stress the strong assumptions commonly made by market participants. We then introduce CCS specifics and highlight how their valuation would require defining cross-currency dynamical models. Finally, a theoretical bootstrap procedure for discounting and forwarding curves, which could be used in presence of a sufficiently rich market, is hinted, underlining which piece of information is proper to each quote typology. A practical curve bootstrapping approach will be described in Section \ref{sec:effectiveDisc}.

\subsection{Pricing FX Swaps in the Market Practice}

According to the pricing formulae listed in Table~\ref{tab:pricingformula} derivative contracts collateralized in different currencies have different prices. On the other hand, even if price differences are present, liquid market instruments, such as FX swaps, are usually quoted without mentioning the currency used for collateralization, since uncertainties are usually hidden in the bid-ask spread quoted market. We investigate the consequence of this approximation.

\subsubsection{Changing the Collateralization Currency}

We consider a domestic investor funding in foreign currencies by means of FX swap contracts collateralized in domestic currency, as we did in the previous section. In this setting, we wish to price a FX swap collateralized in foreign currency and remunerated at the foreign overnight rate $e^f_t$. We can apply equation \eqref{eq:pricingforeign} to obtain
\[
V^{\rm FXswap/f}_t := \Ex{t}{ \left(\frac{\chi_T}{X_t(T;e^f,e)} - 1\right) D(t,T;e^f-b^f(e)+e) }\,,
\]%
where we name $X_t(T;e^f,e)$ the par rate of the contract. We can solve for the par rate to get
\[
X_t(T;e^f,e) = \frac{ \Ex{t}{ \chi_T D(t,T;e^f-b^f(e)+e)} }{ \Ex{t}{D(t,T;e^f-b^f(e)+e)} } = X_t(T;e) \left( 1 + \gamma^\chi_t(e^f,e) \right)
\]%
where the convexity $\gamma^\chi_t(e^f,e)$ of the FX forward rate due to a change in collateral currency is defined as
\[
\gamma^\chi_t(e^f,e) := \frac{ \cov{t}{T;e}{\chi_T}{D(t,T;e^f-b^f(e))} }{ X_t(T;e) \,\ExT{t}{T;e}{D(t,T;e^f-b^f(e))} }\,.
\]%
We notice that, if the above covariance is null, e.g. this occurs when the spread between the basis rate $b^f(e)$ and the foreign overnight rate $e^f$ is a deterministic function of time, then the convexity is zero, and, in turn, the par rate is equal to the forward rate given in equation \eqref{eq:fxforward}.

In general, we cannot estimate the convexity $\gamma^\chi_t(e^f,e)$ from market data, since $V^{\rm FXswap}_t$ and $V^{\rm FXswap/f}_t$ share the same quote on the FX market, and there are not other liquid quotes with this information.

\begin{remark}{\bf (The point of view of the counterparty)}
Market quotes for FX swaps hide another price uncertainty. The price of contract calculated by the investor is not equal to the price calculated by the counterparty if the two parties do not share the same funding strategy. For instance, we can consider a contract with coupons and collateralization in USD traded between a USD-based investor and a EUR-based counterparty. The investor, funding natively in USD, can price the contract by means of equation \eqref{eq:pricing}, the domestic currency being USD, while the counterparty, funding in USD by means of EUR/USD FX swaps, should price the deal with equation \eqref{eq:pricingforeigncouponforeign}, its domestic currency being EUR.
\end{remark}

\subsubsection{Currency Triangulations}
The approach of quoting FX par swap rates regardless of the chosen collateral currency also impacts FX swap triangulations. Let us consider three currencies $\{x,y,z\}$ and the FX swaps between such currencies. Market practice is to quote par swap rates such that 
$$
\frac{X^{x\rightarrow z}_t(T)}{X^{y\rightarrow z}_t(T)}\approx  X^{x\rightarrow y}_t(T)\,. 
$$
By means of the theoretical framework we developed, we want to delimit the validity of such a relationship.

Let us focus on the FX forward rates from currencies $x$, and $y$, to currency $z$. We first assume that the collateral accounts are in currency $z$, and remunerated at the overnight rate $e^z_t$. We name such rates  $X^{x\rightarrow z}_t(T;e^z)$ and $X^{y\rightarrow z}_t(T;e^z)$. If we calculate the ratio between them, we get
\[
\frac{X^{x\rightarrow z}_t(T;e^z)}{X^{y\rightarrow z}_t(T;e^z)}
= \frac{\Ex{t}{\chi^{x\rightarrow z}_T D(t,T;e^z)}}{\Ex{t}{\chi^{y\rightarrow z}_T D(t,T;e^z)}}
=  \chi^{x\rightarrow y}_t \,\frac{P^x_t(T;e^z)}{P^y_t(T;e^z)}
\]%
where the last step comes from the triangulation rules of spot FX market, namely
\[
\chi^{x\rightarrow z}_t = \chi^{x\rightarrow y}_t \chi^{y\rightarrow z}_t\,.
\]%
On the other hand, by equation \eqref{eq:pricingcouponforeign} the FX forward rate from currency $x$ to currency $y$ with collateralization in currency $z$ at $e^z_t$ overnight rate is given by
\[
X^{x\rightarrow y}_t(T;e^z) = \frac{ \ExT{t}{b^y}{ \chi^{x\rightarrow y}_T D(t,u;b^y(e^z))} }{ \ExT{t}{b^y}{ D(t,u;b^y(e^z))} } = \frac{ \Ex{t}{ \chi^{x\rightarrow z}_T D(t,u;e^z)} }{ \chi^{y\rightarrow z}_t \,\ExT{t}{b^y}{ D(t,u;b^y(e^z))} } = \chi^{x\rightarrow y}_t \,\frac{P^x_t(T;e^z)}{P^y_t(T;e^z)}\,.
\]%
Thus, the triangulation rule for FX forward rates holds only if all the rates share the same collateralization currency.
\begin{equation}
X^{x\rightarrow z}_t(T;e^z) = X^{x\rightarrow y}_t(T;e^z) \,X^{y\rightarrow z}_t(T;e^z)\,.
\end{equation}%

The same relationship does not hold if we change the collateralization currency. For instance, we consider the FX forward rates from currency $x$ to currency $y$ with collateralization in currency $y$ at $e^y_t$ overnight rate. We get from equation \eqref{eq:pricingforeigncouponforeign}
\begin{eqnarray*}
X^{x\rightarrow y}_t(T;e^y)
&=& \frac{ \ExT{t}{b^y}{ \chi^{x\rightarrow y}_T D(t,u;e^y)} }{ \ExT{t}{b^y}{ D(t,u;e^y)} } = \frac{ \Ex{t}{ \chi^{x\rightarrow z}_T D(t,u;e^y-b^y(e^z)+e^z)} }{ \Ex{t}{ \chi^{y\rightarrow z}_T D(t,u;e^y-b^y(e^z)+e^z)} } \\
&=& \frac{X^{x\rightarrow z}_t(T;e^z) \left( 1 + \gamma^{\chi^{x\rightarrow z}}_t(e^y,e^z) \right)}{X^{y\rightarrow z}_t(T;e^z) \left( 1 + \gamma^{\chi^{y\rightarrow z}}_t(e^y,e^z) \right)}\\
&=& X^{x\rightarrow y}_t(T;e^z)\left(\frac{ 1 + \gamma^{\chi^{x\rightarrow z}}_t(e^y,e^z)}{ 1 + \gamma^{\chi^{y\rightarrow z}}_t(e^y,e^z)}\right)\,.
\end{eqnarray*}%
If the market approximation 
\begin{equation}
\label{eq:marketapprox}
\gamma^\chi_t(e^y,e^z) \approx 0\,.
\end{equation}%
 is assumed, we obtain that the triangulation rule for FX forward rates is holding again.

\subsection{Cross-Currency Swaps}

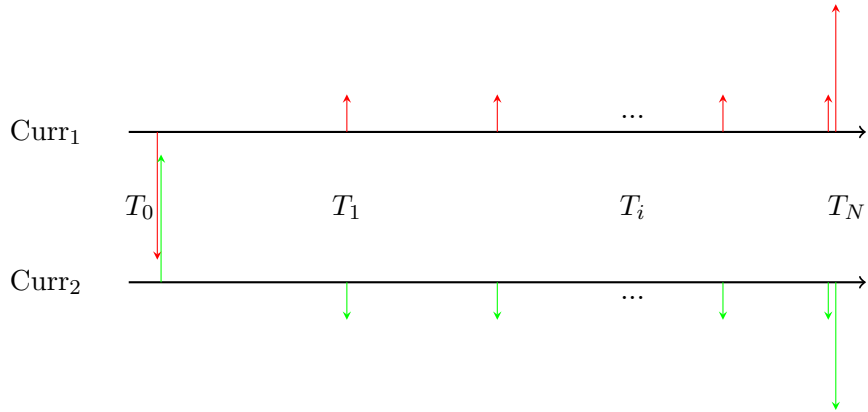
\begin{figure}[t!]
\begin{center}
\begin{tikzpicture}
\node at (-6,1) {Curr${}_1$} ;
\node at (-6,-1) {Curr${}_2$} ;
\draw[->, thick] (-4.9,1) -- (4.9,1);
\draw[->, thick] (-4.9,-1) -- (4.9,-1);
\node at (-4.75,0) {$T_0$};
\node at (-2,0) {$T_1$};
\node at (1.8,0) {$T_i$};
\node at (1.8,1.2) {...};
\node at (4.65,0) {$T_N$};

\draw[red, -stealth] (-4.52,1) -- (-4.52,-0.7);
\draw[red, -stealth] (-2,1)   -- (-2,+1.5);
\draw[red, -stealth] (0,1)    -- (0,1.5);
\draw[red, -stealth] (3,1) -- (3, 1.5);
\draw[red, -stealth] (4.4,1) -- (4.4,1.5);
\draw[red, -stealth] (4.5,1) -- (4.5,2.7);
\node at (1.8,-1.2) {...};
\draw[green, -stealth] (-4.47,-1) -- (-4.47,0.7);
\draw[green, -stealth] (-2,-1)   -- (-2,-1.5);
\draw[green, -stealth] (0,-1)    -- (0,-1.5);
\draw[green, -stealth] (3,-1) -- (3, -1.5);
\draw[green, -stealth] (4.4,-1) -- (4.4,-1.5);
\draw[green, -stealth] (4.5,-1) -- (4.5,-2.7);
\end{tikzpicture}
\end{center}
\caption{\label{fig:CNCCS} A cash flow description of a constant-notional CCS.} 
\end{figure}

FX swaps are quoted with sufficient liquidity only for short maturities, i.e. up to two-four years depending on the considered currency pair. For longer maturities, market participant exchange amounts of currency by means of CCS, which can be thought to as pairs of coupon-paying loans denominated in two different currencies. In the simplest case, depicted in Figure~\ref{fig:CNCCS}, each of the two parties lends to the other an amount of money at the swap start date $T_0$, receives for it (floating) rate interests at dates $T_1,\ldots,T_i,\ldots,T_N$, and gets the notional back at maturity $T_N$. These CCS are called constant-notional because the principal amount used to value interests is established once for all at inception.
For most currency pairs, standardized CCS are structured such that, at inception, notionals are equivalent (using FX rate for $T_0$) and the deal is entered at-par.

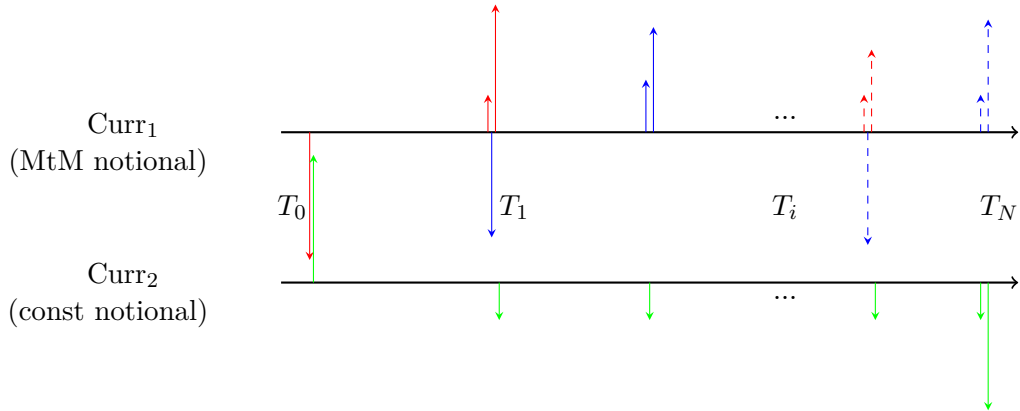
\begin{figure}[t!]
\begin{center}
\begin{tikzpicture}
\node at (-7,1.1) {Curr${}_1$};
\node at (-7.2,0.6) {(MtM notional)};
\node at (-7,-0.9) {Curr${}_2$} ;
\node at (-7.2,-1.4) {(const notional)} ;
\draw[->, thick] (-4.9,1) -- (4.9,1);
\draw[->, thick] (-4.9,-1) -- (4.9,-1);
\node at (-4.75,0) {$T_0$};
\node at (-1.8,0) {$T_1$};
\node at (1.8,0) {$T_i$};
\node at (1.8,1.2) {...};
\node at (4.65,0) {$T_N$};

\draw[red, -stealth] (-4.52,1) -- (-4.52,-0.7);
\draw[red, -stealth] (-2.15,1)   -- (-2.15,+1.5);
\draw[red, -stealth] (-2.05,1)   -- (-2.05,+2.7);
\draw[blue, -stealth] (-2.1,1)   -- (-2.1,-0.4);
\draw[blue, -stealth] (-0.05,1)    -- (-0.05,1.7);
\draw[blue, -stealth] (0.05,1)    -- (0.05,2.4);
\draw[red, dashed, -stealth]  (2.95,1) -- (2.95, 2.1);
\draw[red, dashed, -stealth]  (2.85,1) -- (2.85, 1.5);
\draw[blue, dashed, -stealth] (2.9,1) -- (2.9, -0.5);
\draw[blue, dashed, -stealth] (4.4,1) -- (4.4,1.5);
\draw[blue, dashed, -stealth] (4.5,1) -- (4.5,2.5);
\node at (1.8,-1.2) {...};
\draw[green, -stealth] (-4.47,-1) -- (-4.47,0.7);
\draw[green, -stealth] (-2,-1)   -- (-2,-1.5);
\draw[green, -stealth] (0,-1)    -- (0,-1.5);
\draw[green, -stealth] (3,-1) -- (3, -1.5);
\draw[green, -stealth] (4.4,-1) -- (4.4,-1.5);
\draw[green, -stealth] (4.5,-1) -- (4.5,-2.7);
\end{tikzpicture}
\end{center}
\caption{\label{fig:MtMCCS} A cash flow description of a marked-to-market CCS.} 
\end{figure}

 The FX market mainly quotes marked-to-market (MtM) CCS, which are built by appending a series of at-par single period CCS one after the other. The resulting contract is shown in Figure~\ref{fig:MtMCCS}, and it behaves like a pair of rolling loans, where notionals are exchanged at each interest rate payment date, hence reducing the counterparty risk and the FX risk of the structure. Legs with a MtM notional are dubbed renotioning or resetting legs, to distinguish them from constant-notional ones.

\medskip

\noindent Most of quoted and liquid CCS have the following features:
\begin{itemize}
\item the major currency has a renotioning leg, while a minor currency has a constant-notional leg;
\item the major currency has interests indexed to flat Libor rates, while a minor currency has interests based on Libor rates plus a spread;
\item the spread over a minor currency floaters is chosen such that the CCS is at-par (equilibrium spreads are CCS market quotes);
\item payments occur quarterly
\end{itemize}
As examples of quoted CCS that presents non-standard characteristics, we find the CNH fixed rate \textit{versus} USD floating rate and the non-deliverable CNY fixed rate vs USD floating rate, both of which without renotioning.

\medskip
In order to employ CCS in curves calibration, we need to value their net present value (NPV). For sake of clarity, in the following sections we take the point of view of a domestic investor pricing FX swaps, constant-notional CCS and marked-to-market CCS with collateral posted in domestic currency and remunerated at the same rate $e;$ therefore all foreign flows will be priced by the formula in the third row of Table~\ref{tab:pricingformula}. The extension to different funding strategies is straightforward by means of the results of  Section \ref{sec:pricing}.
\subsubsection{Constant-Notional CCS}
\label{sec:ConstNotCCS} 
Let us consider a constant-notional CCS, where interests are indexed to domestic and foreign Libor rates $L_t^x(T),$ $t$ being the fixing date, $T$ the maturity and $x\in\{d,f\} $. We assume that the domestic market quotes single-currency interest rate swaps with a floating leg indexed to the same Libor rates the CCS domestic leg is indexed to and where the standard collateralization is based on the same collateral rate $e.$ Hence, we define the domestic Libor forward rate $F_t(T_i;e)$ as the forward for the Libor rate $L_{T_{i-1}}(T_i)$  when the collateral is posted in natural currency, i.e. 
\begin{equation}
\label{eq:domesticfwdlibor}
F_t(T_i;e) := \frac{ \Ex{t}{ D(t,T_i;e) L_{T_{i-1}}(T_i) } }{ P_t(T_i;e) } = \ExT{t}{T_i;e}{L_{T_{i-1}}(T_i)}\,.
\end{equation}%
For sake of generality we allow CCS interests to be equal to Libor rates plus a spread, which we name $s$ for the domestic leg and $s^f$ for the foreign leg. According to first and third rows of Table \ref{tab:pricingformula}, we compute the CCS price, expressed in domestic units, as
\begin{equation}
\label{eq:CNCCS}
V^{\rm CCS}_t := V^{\rm CCS/d}_t - V^{\rm CCS/f}_t
\end{equation}%
where the net present values of the domestic and foreign legs are given by
\begin{eqnarray}
\label{eq:CNCCSd}
V^{\rm CCS/d}_t &:=& N \left( -P_t(T_0;e) + \sum_{i=1}^N \tau_i \left(F_t(T_i;e) + s \right) P_t(T_i;e) + P_t(T_N;e) \right) \\
\label{eq:CNCCSf}
V^{\rm CCS/f}_t &:=& \chi_t N^f \left( -P^f_t(T_0;e) + \sum_{i=1}^N \tau_i \left( F^f_t(T_i;e) + s^f \right) P^f_t(T_i;e) + P^f_t(T_N;e) \right)
\end{eqnarray}%
and where $N$ and $N^f$ stand respectively for the notionals of the domestic and foreign legs, $\tau_i$ is the year fraction calculated between $T_{i-1}$ and $T_i$. Most importantly, in valuing the net present value of the foreign leg  we introduced the foreign basis forward Libor rates $F^f_t(T_i;e)$ observed under domestic collateralization as
\begin{equation}
\label{eq:foreignfwdlibor}
F^f_t(T_i;e) := \frac{\ExT{t}{b}{ D(t,T_i;b^f(e)) L^f_{T_{i-1}}(T_i)} }{ P^f_t(T_i;e) } = \ExT{t}{T_i;b}{L^f_{T_{i-1}}(T_i)}\,.
\end{equation}%

\begin{remark}{\bf (Forward Libor rates depend on collateral and funding strategies)}
\label{rem:basisFwdLibor}
Single-currency products quoted on the foreign money market are usually collateralized in the same foreign currency. For instance, for interest-rate swaps quoted in these markets we can bootstrap foreign basis forward Libor rates $F^f_t(T_i;e^f)$ observed under foreign collateralization. We can introduce them by means of Equation~\eqref{eq:pricingforeigncouponforeign} as given by
\[
F^f_t(T_i;e^f) := \frac{\ExT{t}{b}{ D(t,T_i;e^f) L^f_{T_{i-1}}(T_i)} }{ \ExT{t}{b}{ D(t,T_i;e^f) } } \,.
\]
Yet, since we observe market dislocations between on-shore and off-shore market players, we cannot assume that the above pricing formula, which is consistent with a foreign funding strategy based on FX swaps and is used by domestic investors, gives the same prices calculated by foreign investors which may recur to other funding sources. Such foreign investors define the forward Libor rates ${\hat F}^f_t(T_i;e^f)$ by means of an analogous of Equation~\eqref{eq:domesticfwdlibor}. Since $F^f_t(T_i;e)$ and $F^f_t(T_i;e^f)$ rates are difficult to bootstrap from market quotes, in the following, when dealing with effective curve bootstrapping procedures, we will use ${\hat F}^f_t(T_i;e^f)$ rates as a viable proxy.
\end{remark}


%
\subsubsection{Marked-to-Market Contributions}
\label{sec:MtMLegs}

We then focus on the pricing of CCS with marked-to-market feature. The resetting of the notional creates an asymmetry between the two legs, so that different pricing formulae will be needed according to the leg on which the marking-to-market operates.

If the renotioning leg is the domestic one, we get 
\begin{equation}
\label{eq:MtMCCS_2}
V^{\rm CCS}_t := V^{\rm MtMCCS/d}_t - V^{\rm CCS/f}_t
\end{equation}%
where the net present value of the MtM leg is given by
\begin{eqnarray}
\label{eq:VCCSd_mtm}
V^{\rm MtMCCS/d}_t
&=& N^f \,\sum_{i=1}^{N} P_t(T_i;e)\,\ExT{t}{T_i;e}{\chi_{T_{i-1}}\left(1+\tau_i \left(L_{T_{i-1}}(T_i)+s\right)\right)} \nonumber\\
&-& N^f \,\sum_{i=1}^{N} P_t(T_{i-1};e)\,\ExT{t}{T_{i-1};e}{\chi_{T_{i-1}}} 
\end{eqnarray}
while the constant-notional leg is defined as in Equation~\eqref{eq:CNCCSf}.

In the other case, where the foreign leg has a renotioning feature, we have
\begin{equation}
\label{eq:MtMCCS_1}
V^{\rm CCS}_t := V^{\rm CCS/d}_t - V^{\rm MtMCCS/f}_t
\end{equation}%
where the net present value of the MtM leg reads
\begin{eqnarray}
\label{eq:VCCSf_mtm}
V^{\rm MtMCCS/f}_t
&=& \chi_t N \,\sum_{i=1}^{N} P^f_t(T_i;e)\ExT{t}{T_i;b}{\frac{1}{\chi_{T_{i-1}}}\left(1+\tau_i \left(L^f_{T_{i-1}}(T_i)+s^f\right)\right)} \nonumber\\
&-& \chi_t N \,\sum_{i=1}^{N} P^f_t(T_{i-1};e)\ExT{t}{T_{i-1};b}{\frac{1}{\chi_{T_{i-1}}}}
\end{eqnarray}%
while the constant-notional leg is defined as in Equation~\eqref{eq:CNCCSd}.

These two pricing formulae share the same structure. The contribution of the first summation represents the redemption of the lending which occurrs at the end of each period plus the payment of matured interests, while the second summation corresponds to the lending of an amount of currency at each coupon period start date, such as to be at par with the constant-notional of the other leg. Let us now separately analyze the structure of the MtM contributions.

\subsubsection*{Domestic Marked-to-Market Leg}

We start by analyzing the domestic MtM leg. The second summation term in equation \eqref{eq:VCCSd_mtm}, where the exchange rate read at $T_{i-1}$ is immediately paid, corresponds to the flow of an FX swap, and it is simply given by
\[
\sum_{i=1}^{N}P_t(T_{i-1};e)\ExT{t}{T_{i-1};e}{\chi_{T_{i-1}}} = \chi_t\sum_{i=1}^{N} P^f_t(T_{i-1};e)\,.
\]%

On the other hand the first summation, over interests and notional repayments, is more tricky, because it involves two terms linked to the correlation structure of discounting, forwarding and exchange rates. In particular we need to evaluate a FX forward with delayed payment, namely
\begin{equation}\label{eq:domM2M_1}
\ExT{t}{T_i;e}{\chi_{T_{i-1}}} = \ExT{t}{T_{i};e}{X_{T_{i-1}}(T_{i-1};e)}
\end{equation}%
and a floating domestic payment with stochastic notional
\begin{equation}\label{eq:domM2M_2}
\ExT{t}{T_i;e}{ \chi_{T_{i-1}}L_{T_{i-1}}(T_i)} = \ExT{t}{T_i;e}{X_{T_{i-1}}(T_{i-1};e)F_{T_{i-1}}(T_i;e)}\,.
\end{equation}%

We wrote these contributions in terms of the forward exchange rate $X_{t}(T_{i-1};e)$ and of the forward Libor rate $F_{t}(T_i;e)$, which are martingales under the terminal measures $\mathbb{Q}^{T_{i-1};e}$ and $\mathbb{Q}^{T_{i};e}$, respectively. These two measures are linked by standard Radon-Nikodym derivative
\begin{equation}
\label{eq:dQidQim1}
Z_t(T_{i-1},T_i;e)
:= \left.\frac{d \mathbb{Q}^{T_i;e}}{d \mathbb{Q}^{T_{i-1};e}}\right|_t
 = \frac{P_t(T_i;e)}{P_t(T_{i-1};e)} \,\frac{P_0(T_{i-1};e)}{P_0(T_i;e)}\,.
\end{equation}

As a consequence, the first term must be valued taking into account the correlation between the forward exchange rate and the collateral curve (pure change of measure effect), while the second also need to incorporate the correlation with forward Libors (change of measure plus covariation). Any estimate of these contributions requires defining joint distribution with covariation effects, see Appendix \ref{sec:DomM2MEstimate} for a simple proposal. We notice that the FX delayed payment terms would be present even in the case of a fixed rate MtM CCS.

\subsubsection*{Foreign Marked-to-Market Leg}

We focus on the foreign MtM leg, whose tractation is totally analogous to the one carried on for the domestic case. The term bound to the lending of a marked-to-market foreign notional is trivial, because it states that we are lending one unit of domestic currency at each date $T_0,\ldots,T_{N-1}$ and we get, by means of Eq.\eqref{eq:fromQtoQb}, 
\[
\chi_t \,\sum_{i=1}^{N}P^f_t(T_{i-1};e) \,\ExT{t}{T_{i-1};b}{\frac{1}{\chi_{T_{i-1}}}} = \sum_{i=1}^{N} P_t(T_{i-1};e)\,.
\]%
Analogously to the domestic MtM case we then write the terms related to the payment of interests and notional redemptions relying on forward exchange rates and forward Libor rates. We have a FX forward with delayed payment
\begin{equation}\label{eq:forM2M_1}
\ExT{t}{T_i;b}{\frac{1}{\chi_{T_{i-1}}}} = \ExT{t}{T_i;b}{\frac{1}{X_{T_{i-1}}(T_{i-1};e)}}
\end{equation}%
and a floating foreign payment with stochastic notional
\begin{equation}\label{eq:forM2M_2}
\ExT{t}{T_i;b}{\frac{L^f_{T_{i-1}}(T_i)}{\chi_{T_{i-1}}}} = \ExT{t}{T_i;b}{\frac{F^f_{T_{i-1}}(T_i;e)}{X_{T_{i-1}}(T_{i-1})}} \,.
\end{equation}%

In this case the exchange rate to be used for notional purposes is the reverse rate $1/\chi_t$ converting domestic units into foreign one. Both expectations involve the basis forward measure $\mathbb{Q}^{T_{i};b},$ under which the basis forward Libor rate $F^f_t(T_{i};e)$ is a martingale. The forward exchange rate $X_{t}(T_{i-1};e),$ as seen, is a  martingale under the terminal measure $\mathbb{Q}^{T_{i-1};e}, $ which is connected to $\mathbb{Q}^{T_{i};b}$ by means of the Radon-Nikodym derivative 
\begin{equation}
\label{eq:dQibasisdQim1}
Z^f_t(T_{i-1},T_i;e)
:= \left.\frac{d \mathbb{Q}^{T_i;b}}{d \mathbb{Q}^{T_{i-1};e}}\right|_t
 = \frac{X_{t}(T_{i-1};e) P^f_t(T_i;e)}{P^f_t(T_{i-1};e)}\cdot\frac{P^f_0(T_{i-1};e)}{X_{0}(T_{i-1};e) P^f_0(T_i;e)}\,.
\end{equation}%

Once again, the first expectation involves a change of measure contribution, and depends on the correlation between $X_{t}(T_{i-1};e)$ and $Z^{i-1,i}_t(e;b),$ while the second also depends on the interplay with basis forward Libor rates. Analogously to the MtM domestic leg case, we can estimate these terms by means of very simple market model, as detailed in Appendix \ref{sec:forM2MEstimate}.

\subsection{Bootstrapping market information}
In the present section we tie together all the analysis brought on so far and quickly sketch a hypothetical bootstrapping procedure to infer  basis discount factors, basis forward Libor rates, as well as correlation terms. We assume here that the market quotes, for a set of maturities, FX swaps, constant-notional CCS, and MtM CCS both with fixed and floating rate interests. This procedure only constitute a theoretical case study and we will discuss in Section \ref{sec:effectiveDisc} the approximations allowing to extract the relevant information from available quotes.

\medskip
The first step consists in applying Equation~\eqref{eq:switch} to a set of domestic-foreign FX swap with increasing maturities $T_1,T_2,\ldots,T_N,$ such as to derive the basis discount factors
\footnote{Very often, for sake of interpolation purposes, the curve $P_t^f(T;e)$ will be thought of as well described by a reference foreign discount curve $P^f_t(T;e^f_{ref})$ times a zero coupon spread $Z$ as
\begin{equation}
P_t^f(T_i;e) := P^f_t(T;e^f_{\rm ref}) \,e^{-Z(T;e)(T-t)}\,.
\end{equation}  %
The most natural choice for the reference curve is the one deduced by a foreign investor from market products indexed to the main collateral rate for the foreign market. This latter very often coincides with the rate for overnight unsecured deposits. By choosing the spread curve approach, one could bootstrap, for each $T_1,\ldots,T_N, $ zero-spread correctors $Z(T_1),\ldots,Z(T_N).$} $P_t^f(T_i;e).$ These basis zero-coupon bonds constitute the fundamental pillars of a basis discounting curve used to discount foreign flows collateralized in domestic currency. 

Second, we consider constant-notional CCS and apply formulae of Sec.\ref{sec:ConstNotCCS} to deduce basis forward Libor rates $F^f_t(T_i;e)$ for a set of maturities $T_i$ hence building a basis forwarding curve.

Marked-to-market CCS with domestic renotioning leg and fixed interests let us infer forward FX rates with deferred payments (Eq.\eqref{eq:domM2M_1}), which incorporate the correlation between domestic discounts and FX forwards, while CCS with domestic renotioning leg and floating interests are used to deduce the terms involving correlations between FX forwards and domestic Libor forwards (Eq.\eqref{eq:domM2M_2}).

Analogously, Marked-to-market CCS with foreign renotioning leg and fixed interests let us value inverse forward FX rates with deferred payments (Eq.\eqref{eq:forM2M_1}), which incorporate the correlation between basis discounts and FX forwards, while CCS with foreign renotioning leg and floating interests are helpful to estimate correlations between FX forwards and foreign  basis Libor forwards (Eq.\eqref{eq:domM2M_2}). 

In the following section we discuss how to get a practical curve bootstrapping procedure by approximating the pricing formula of MtM CCS. We discuss a market approximation similar to the one used to analyze currency triangulations in Section~\ref{sec:pricing}.

\section{Effective Discounting Curve Approach}
\label{sec:effectiveDisc}
As said before, the only FX quotes which are actively traded are FX swaps with short to mid maturities together with MtM CCS with flat floating interests and renotioning  for the major currency leg versus floating interests plus spread and constant notional on the minor currency leg. The absence of a set of quotes allowing a sequential bootstrap of the foreign basis discounting curve, of the foreign basis forwarding curve and of correlations driving MtM corrections, forces market players need to find some approximations.  
A very common approach to take into account the information embedded in market quotes, and to quickly price CCS, consists in avoiding direct modelling of the dependencies among all the components of the swap price formula, and valuing net present values by means of an effective discounting curve approach, see for example \cite{Fries2012}.

The most natural way for an investor to achieve this result is to give relevance to its own domestic currency and price the CCS by means of four curves:
\begin{itemize}
\item a domestic discounting curve, which is the same curve linked to the domestic collateral rate;
\item a domestic forwarding curve, which is the same curve obtained from single-currency standard floaters quoted in the domestic money market;
\item a foreign currency forwarding curve, which is the same curve obtained from single-currency standard floaters quoted in the foreign currency money market;
\item an implied foreign currency discounting curve, bootstrapped such as market CCS are repriced at par.
\end{itemize}

By means of this procedure we choose to use unadjusted foreign forwards, as if they were paid and collateralized in their own currency, and incorporate all the corrections discussed above into an implied foreign currency discounting curve. In this way we deduce an implied foreign currency curve that does not correspond to the basis foreign currency curve given in equation \eqref{eq:basiszcbond}, and used in the previous section to present the theoretical curve bootstrapping procedure, unless some approximations hold. 
 
\subsection{Bootstrapping Curves in the FX Market}

The short end of the implied curve could be straightforwardly stripped by FX swaps by means of Eq.\eqref{eq:switch}. Thus, if we call $T_c$ the longest maturity for which we can find on the market liquid quotes of FX swaps, we can write
\[
P^{f,{\rm impl}}_t(T;e) := \frac{X_t(T;e)}{\chi_t}P_t(T;e)
\;,\quad
T\le T_c
\] %
where $X_t(T;e)$ is given by the market. 

Then, in order to cover the mid-long part of FX curves ($T>T_c$), we need to develop simplified pricing formulae for CCS to be used for bootstrapping purposes. The effective curve approach consists in disregarding all of the contributions that would require a dynamical model and in re-writing in a simple way the net present values given by equations \eqref{eq:CNCCSf}, \eqref{eq:VCCSd_mtm}, \eqref{eq:VCCSf_mtm} in terms of implied basis foreign zero-coupon bonds $P^{f,{\rm impl}}_t(T;e).$ These latters will be calibrated such as to ensure that a set of  relevant market instruments is priced at-par. Equation \eqref{eq:CNCCSd}, wich only involves domestic flows and domestic collateral, is unchanged.

Let us begin by analyzing the MtM domestic leg of a CCS. Its NPV can be cast in the form
\begin{equation}
\label{eq:VCCSd_mtm_approx}
V^{\rm MtMCCS/d}_t
= \sum_{i=1}^{N} N^{\rm impl}_{i-1} \left( - P_t(T_{i-1};e) + P_t(T_i;e) \left( 1 + \tau_i (F_t(T_i;e) + s ) \right) \right)
\end{equation}
by introducing the coupon-dependent notionals
\[
N^{\rm impl}_i := N^f X^{\rm impl}_t(T_i;e)
\]
based on the implied forward exchange rates
\[
X^{\rm impl}_t(T_i;e):=\frac{\chi_tP^{f,{\rm impl}}_t(T_i;e)}{P_t(T_i;e)}\,.
\]%
As for the foreign leg, in presence of constant notional (Eq.\eqref{eq:CNCCSf}), we set
\begin{equation}
\label{eq:CNCCSf_effective}
V^{\rm CCS/f}_t =  \chi_t N^f \left( -P^{f,{\rm impl}}_t(T_0;e) + \sum_{i=1}^N \tau_i \left( {\hat F}^f_t(T_i;e^f) + s^f \right) P^{f,{\rm impl}}_t(T_i;e) + P^{f,{\rm impl}}_t(T_N;e) \right)\,,
\end{equation}%
while if the leg is marked-to-market (Eq.\eqref{eq:VCCSf_mtm}), we write
\begin{equation}
\label{eq:VCCSf_mtm_approx_1}
V^{\rm MtMCCS/f}_t
= \chi_t \sum_{i=1}^{N} N^{f,{\rm impl}}_{i-1} \left( - P^{f,{\rm impl}}_t(T_{i-1};e) + P^{f,{\rm impl}}_t(T_i;e) \left(1+\tau_i ({\hat F}^f_t(T_i;e^f) + s^f) \right) \right)
\end{equation}%
where we defined the maturity-dependent notionals 
\[
N^{f,{\rm impl}}_i := N \frac{1}{X^{\rm impl}_t(T_i;e)}\,.
\]%
and, in the formulae related to foreign leg, we replaced basis foreign forward Libor rates with the foreign forward Libor rates ${\hat F}^f_t(T_i;e^f)$ which are the rates bootstrapped by a foreign investor by means of its own money market quotes with the analogous of Equation~\eqref{eq:domesticfwdlibor} discussed in Remark~\ref{rem:basisFwdLibor}. We use these rates because we are able to bootstrap them from market quotes.

\medskip
The rationale behind this definition of implied curve is removing from the equation all the terms depending on dynamical parameters which cannot be bootstrapped by independent market quotes. In the following sections we highlight the terms we have approximated to understand the hypothesis under which the implied curve $P^{f,{\rm impl}}_t(T;e)$ can be identified with the basis curve $P^f_t(T;e)$.

\subsubsection{Approximating a MtM Domestic Leg}

We consider Eq.\eqref{eq:VCCSd_mtm} and, neglecting correlations between FX spot rate and interest rate risk factors, get 
\[
\ExT{t}{T_i}{\chi_{T_{i-1}}} \approx X_t(T_{i-1};e)
\;,\quad
\ExT{t}{T_i}{\chi_{T_{i-1}}L_{T_{i-1}}(T_i)} \approx X_t(T_{i-1};e) F_t(T_i;e)\,.
\]%
The assumption 
\begin{equation}
\label{eq:proxyX}
X^{\rm impl}_t(T_i;e)\approx X_t(T_i;e)\,,
\end{equation}
leads to Eq.\eqref{eq:VCCSd_mtm_approx} and to the identification $P^{f}_t(T_{i};e)\approx P^{f,{\rm impl}}_t(T_{i};e).$
\subsubsection{Approximating a Constant-Notional Foreign Leg}

Let us focus on Eq.\eqref{eq:CNCCSf}. Since the foreign basis forward Libor rates $F^f_t(T_i;e)$ in such equation cannot be deduced from independent market quotes, a possible solution is to replace the forward Libor rate by means of the foreign forward Libor rates as given by the single-currency foreign money market, namely
\begin{equation}
\label{eq:proxyFf}
F^f_t(T_i;e)\approx {\hat F}^f_t(T_i;e^f)\,.
\end{equation}
This choice naturally suggests to identify $P^{f}_t(T_{i};e)\approx P^{f,{\rm impl}}_t(T_{i};e)$ and to price this leg by means of Eq.\eqref{eq:CNCCSf_effective}.  

\subsubsection{Approximating a MtM Foreign Leg}

Finally, we consider Eq.\eqref{eq:VCCSd_mtm}, disregard correlations between FX spot rate and interest rate risk factors to approximate
\[
\ExT{t}{T_i;b}{\frac{1}{\chi_{T_{i-1}}}} \approx \frac{1}{X_t(T_{i-1};e)}
\;,\quad
\ExT{t}{T_i;b}{\frac{L_{T_{i-1}}(T_i)}{\chi_{T_{i-1}}}} \approx \frac{F^f_t(T_i;e)}{X_t(T_{i-1};e)}
\]%
and get a simple formula for the net present value of the foreign leg of a MTM CCS as
\[
V^{\rm MtMCCS/f}_t
\approx \chi_t \sum_{i=1}^{N} N^f_{i-1} \left( - P^f_t(T_{i-1};e) + P^f_t(T_i;e) \left(1+\tau_i (F^f_t(T_i;e) + s^f) \right) \right)\,.
\]%

Under the approximations of Eqs.\eqref{eq:proxyX} and \eqref{eq:proxyFf} once again we can inteerprete the implied discounts $P^{f,{\rm impl}}_t(T_{i};e)$ as proxy for the true basis discount $P^{f}_t(T_{i};e).$ 

\subsection{Numerical Examples}

In this section we want to provide numerical examples of the application of the effective discounting curve approach just described. First, we consider the currency pair USD/EUR, bootstrap implied foreign discount curves from MtM CCS (renotioning is on the USD leg) and use them to price constant-notional CCS. Second, we investigate the case of an investor wanting to price a EUR/HKD CCS by triangulation through EUR/USD and HKD/USD. 

\subsubsection{Constant-Notional \textit{versus} MtM CCS }

\begin{table}[htp]
  \caption{\label{t:mktTripletQuotes}{\small
  Market par-spreads (\% units) for the pair USD/EUR for different CCS maturities (in months). Data of September, the 6th, 2013. USD leg is marked-to-market, EUR leg pays floating rate plus spread. }}
  \begin{center}
   \begin{tabular}{|c|c|}\hline
   \rule[-1ex]{0ex}{4ex}
   Maturity&  USD/EUR  \\\hline\hline
   \rule[-0.5ex]{0ex}{3ex}
   1y    &-0.1450 \\
   18m   & -0.1850 \\
   2y    &-0.2050 \\
   3y    &-0.2375 \\
   4y    &-0.2550 \\
   5y    &-0.2650 \\
   7y    &-0.2675 \\
   10y   &-0.2625 \\
   15y   &-0.2475 \\
   20y   &-0.2325\\
   30y   &-0.2050\\\hline
   \end{tabular}%
  \end{center}
\end{table}%
%

Let us consider market data of the 6th of September 2013 for the EUR/USD CCS reported in Table \ref{t:mktTripletQuotes}. First of all, market par-spreads were interpolated with a monotone cubic spline such as to get a smooth annual grid of CCS equilibrium spreads. Then, starting from USD-OIS, USD-3M, EUR-OIS, EUR-3M curves\footnote{The EUR 3M curve is bootstrapped with exogenous discounting at EUR-OIS, while the GBP 3M and USD-3M curves are self-discounted.}, we bootstrapped effective curves from CCS quotes assuming the point of view of both a EUR and a USD investor. We first pretend to be an EUR based investor with EUR collateral and bootstrap a USD implied curve. Then we consider the same quotes from the point of view of a USD investor with collateral posting in USD and calibrate a EUR basis implied curve.

\begin{figure}[htp]
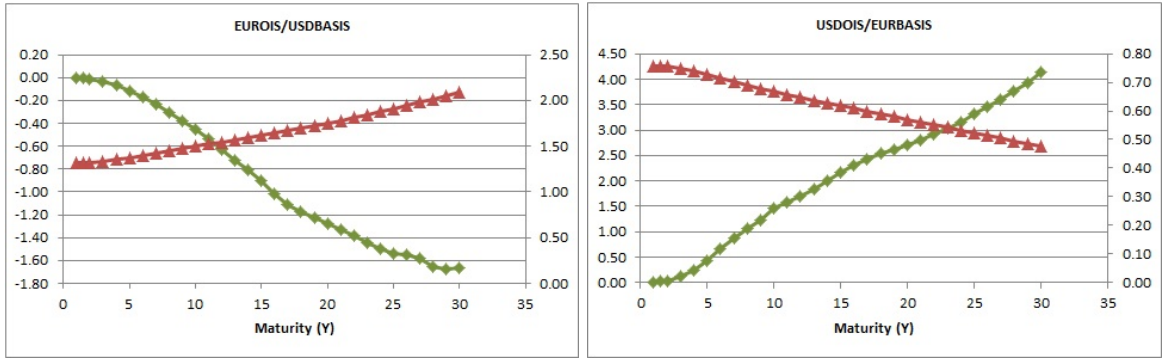

\begin{center}
\includegraphics[width=0.49\textwidth]{figure/EUROIS-USDBS-M2MvsNONE.jpg} 
\includegraphics[width=0.49\textwidth]{figure/USDOIS-EURBS-M2MvsNONE.jpg}\\
\caption{\small{Par-spread differences in bps between marked-to-market and constant-notional CCS (green diamonds,left scale). Implied FX forwards are also plotted (red triangles, right scale).}}\label{f:parSprM2MvsNone}
\end{center}
\end{figure}

Differences in basis points between MtM and constant-notional CCS equilibrium spreads are plotted in Figure \ref{f:parSprM2MvsNone} as a function of the CCS maturity. In the same plots we also report, on secondary axis, the forward FX rates implied by the CCS bootstrap. Even if forward FX rates are generally consistent when we switch from the EUR-based approach to the the USD-based point of view, constant-notional equilibrium spreads may differ between the first and the second configuration.  Unfortunately, it is not possible to estimate the error that is done when we associate the very same CCS spreads to a swap collateralized in first or second currency. The differences in par-spreads MtM \textit{vs} constant-notional may be attributed to the use of a unique implied curve to incorporate all of the effects.

\subsection{Currency Triplet Consistency}
\label{sec:triplet}

We consider a triplet of currencies $\{x,y,z\}$, with $x$ prevailing over $y$, and $y$ over $z$. The market usually does not quote CCS par-spreads for all currency pairs. For instance, we consider the currency triplet \{USD, EUR, HKD\}. In this case the Hong-Kong dollar (HKD) is liquidely swapped only against USD (with implicit collateral in USD), but not against EUR, since the FX market does not quote EUR/HKD CCS. Thus, we are not able to directly deduce a HKD basis curve to discount HKD flows when collateral is posted in EUR. In order to overcome this lack of information, we can build two reasonable triangulation scheme.
\begin{itemize}
\item[(a)] We consider the USD/EUR MtM CCS, with implicit USD collateralization. If we assume that the quotes are the same when collateral is posted in EUR, we can deduce a basis curve used for USD flows collateralized in EUR. Then, we consider the USD/HKD MtM CCS and we equally assume that the quotes do not (significantly) change if the collateral is posted in EUR. We discount the USD leg of this CCS with the basis curve previously built and infer an implied HKD discounting curve to be used when collateral is posted in EUR.
\item[(b)] Starting from USD/EUR and USD/HKD MtM CCS, whose implicit collateral is in USD, we build two implied curves, one to discount EUR flows collateralized in USD, and one for HKD flows collateralized in USD. Then we structure a EUR/HKD MtM CCS where a plain EURIBOR leg is paid against a HKDIBOR plus spread leg; the collateral of this CCS is assumed to be posted in USD, such that we use the implied EUR and HKD curves we have just set up. We require this CCS to be entered at par and find equilibrium spreads for the relevant maturities, under USD collaterilaziation.  To conclude, we assume those spreads do not change (significantly) if the collateral is posted in EUR and use them in a EUR/HKD MtM CCS where EUR leg is discounted at EUR collateral rate to deduce an implied HKD discounting curve.
\end{itemize}

From a computational point of view, the first approach is more efficient, as it requires the estimation of two basis curves (the USD with collateral in EUR and the HKD with collateral in EUR) while the second method requires to build three implied curves (the EUR with collateral in USD, the HKD with collateral in USD and the HKD with collateral in EUR).

In order to check the consistency of the two approaches, that is the meaningfulness of the assumption of independence of par-spreads from the collateral choice, we focus on implied, unquoted, EUR/HKD par spreads. In the (a) framework we value par-spreads with EUR collateral by discounting the EUR leg with EUR-overnight curve and the HKD leg with the HKD implied curve for EUR collateral. In case (b), on the other side, we directly employ the EUR and HKD implied curves against USD getting par-spreads under USD collateralization. We see from the comparison results reported in Table~\ref{t:incompleteTriangulation} that the assumption seems to be verified at a level of accuracy which is compatible with bid-ask spreads for non-quoted currency pairs.

\begin{table}[htp]
  \caption{\label{t:incompleteTriangulation}
  EUR/HKD par-spreads for EUR (a) and USD (b) collateral, and their difference. Spreads in percentage units, differences in basis points.}
  \begin{center}
   \begin{tabular}{|c|ccc|}\hline
   \rule[-1ex]{0ex}{4ex}
   CCS Mat(m) & (a) & (b) & diff \\\hline\hline
   \rule[-0.5ex]{0ex}{3ex}
   12    & 0.0349 & 0.0346 & -0.02 \\
   18    & 0.0680 & 0.0675 & -0.05 \\
   24    & 0.1035 & 0.1025 & -0.10 \\
   36    & 0.1264 & 0.1243 & -0.21 \\
   48    & 0.1476 & 0.1440 & -0.37 \\
   60    & 0.1611 & 0.1568 & -0.44 \\
   84    & 0.1657 & 0.1583 & -0.74 \\
   120   & 0.1738 & 0.1639 & -0.99 \\
   144   & 0.1534 & 0.1381 & -1.53 \\
   180   & 0.1107 & 0.0965 & -1.42 \\\hline
   \end{tabular}%
  \end{center}
\end{table}%

\section{Conclusions and Further Developments}
\label{sec:conclusion}

In this paper we presented a general derivation of the arbitrage-free pricing framework for multiple currency products. We extended the classical approach to include the latest results on collateral and funding costs of \cite{Perini2011} and \cite{Crepey2011}. We showed the impact that the policy adopted to fund in foreing currency has on derivative pricing. In this way we were able to price contracts with cash flows and/or collateral accounts expressed in foreign currencies inclusive of funding costs originating from dislocations in the FX market. 

Then, we applied these results to price CCS to understand how to implement a feasible curve bootstrap procedure. We presented the main practical problems arising from the way the market is quoting liquid instruments: uncertainties about collateral currencies and renotioning features. We discussed the theoretical requirements to implement curve bootstrapping, and the approximations usually taken to practically implement the procedure.

We think that this paper, summarizing and extending under a coherent framework the contribution coming from the practitioner literature, and linking them to the research made in the economics literature on market dislocations, could be used as a starting point for the formulation of specific multiple-currency dynamical models able to include the new features of the FX and money markets appeared after the crisis.

%

\newpage

\appendix

\section{Default-Free Pricing Framework with Collateralization}
\label{sec:appendix}

The margining procedure requires a flow exchange on a scheduled basis to feed the collateral account (usually day-by-day) and to compensate the collateral giver with collateral accrued interests (usually monthly). We deal with the impact of collateralization in pricing equations by following the approach of \cite{BrigoCapponiPallaviciniPapatheodorou} and \cite{Perini2011} where the margining procedure is described as an additional set of coupons (or dividends). Here, we are not interested in a general pricing formula, but we wish to focus on contracts traded under a collateralization agreement which effectively removes counterparty credit risks. Under this assumption we can disregard any default event. In the following we present the pricing equations by following the formalism of \cite{DuffieDAPT}.

\subsection{Derivative Contracts in the Market Practice}

In the interbank market, derivative contracts are traded along with insurances to protect from default events. A practice widely spread after the crisis. An amount of cash or high quality assets is usually posted on a prefixed schedule to the counterparty to match the marked-to-market value of the position. The assets used as insurance are known as collaterals or margins. An introduction to counterparty credit risk and collateralization can be found in \cite{BrigoMoriniPallavicini2012}.

Collaterals are stored in a collateral account $C_t$. How to manage the collateral account during the life of the contract (margining procedure) and what happens on default of one of the counterparties (close-out rules) is regulated by a bilateral agreement documented by ISDA, known as Credit Support Annex (CSA). In particular, the agreement regulates the possibility of re-hypothecating the collateral assets, namely to use them for funding purposes.

Here, we present a stylized description of the margining procedure. We do not consider default events, since we are interested only in the so-called perfect collateralization approximation, which we present in the following section. Under such approximation we assume that the margining procedure is always able to remove all credit risk. For a derivation of pricing equations in presence of credit risk along with funding costs, see \cite{BurgardKjaer2011a,BurgardKjaer2013}, \cite{Perini2011,Perini2012}, or \cite{Crepey2011,Crepey2012a,Crepey2012b}.

We consider a probability space $(\Omega,{\cal A},\mathbb{P})$ endowed with the filtration ${\cal F} := ({\cal F}_t)_{t\ge0}$, where $\mathbb{P}$ is the physical probability measure. We assume that the market is quoting a set of $n$ securities, whose prices are described by the processes $\{V^1_t,\ldots,V^n_t\}$.

We introduce a collateral account $C^k_t$ for each security $V^k_t$. In the practice all the securities traded with the same counterparty under the same CSA are gathered within the same collateral netting set, so that a single collateral account is shared among such securities. In this case we can introduce the price process for the whole security set, and we can proceed as in the following.

Without loss of generality, we consider the assets are posted by the investor if $C^k_t<0$, and by the counterparty if $C^k_t>0$. The bilateral agreement requires that the collateral account must be remunerated at a particular rate level $c^k_t$ by the party receiving collateral assets. The collateral rate is a contractual rate, and, in principle, it has no relationship with the risk-free rate. Most derivatives are bilaterally collateralized on a daily basis at overnight rate.

For each security $V^k_t$ we introduce the cumulated dividend process $D^k_t$ which includes the contractual coupons $\{\gamma^k_1,\ldots,\gamma^k_N\}$, and the cash flows required by the margining procedures
\[
D^k_t := \pi^k_t + C^k_t - \int_0^t c^k_u C^k_u \,du
\;,\quad
\pi^k_t := \sum_{i=1}^N \gamma^k_i \ind{T_i \le t}\,.
\]%
We define also the gain process, namely profits and losses achieved by holding the securities, as given by
\[
G^k_t := V^k_t + D^k_t - C^k_t\,.
\]%
We subtract the collateral account from profits and losses, since the collateral taker withdraw the collateral assets on contract termination.

We assume that for cash-lending and borrowing operations traders can access a risk-free Treasury Bank Account $B_t$ accruing at rate $r_t$.
\[
B_t := \exp\left\{ \int_0^t r_u \,du \right\}\,.
\]%

We use the Treasury bank account as a numeraire, so that we rescale price, collateral and cumulated dividend processes by its value. We extend the discussion of \cite{DuffieDAPT} to include collaterals, and we define the rescaled quantities as given by
\[
{\bar V}^k_t := \frac{V^k_t}{B_t}
\;,\quad
{\bar C}^k_t := \frac{C^k_t}{B_t}
\;,\quad
{\bar D}^k_t := D^k_0 + \int_0^t \frac{dD^k_u}{B_u}
\]%
which allow us to write the rescaled gain process as
\[
{\bar G}^k_t := {\bar V}^k_t + {\bar D}^k_t - {\bar C}^k_t\,.
\]%

Then, to remove arbitrages, we assume the existence of a measure $\mathbb{Q}$ equivalent to the physical measure $\mathbb{P}$, such that the rescaled gain processes ${\bar G}^k_t$ are martingales, so that we can write
\begin{equation}
\label{eq:securitypricing}
V^k_t = B_t \,\Ex{t}{\frac{V^k_T}{B_T} + \int_t^T \frac{d\pi^k_u}{B_u} + \int_t^T \frac{C^k_u}{B_u} (r_u - c^k_u) \,du }
\end{equation}%
where the expectation is taken under the measure $\mathbb{Q}$ and conditioned to ${\cal F}_t$.

\subsection{Perferct Collateralization}

The pricing equation \eqref{eq:securitypricing} obtained in previous subsection is justified if we are able to find a proper margining procedure able to remove any losses on the default event of the investor or the counterparty.

In case of a continuous price process $V^k_t$, namely if we can disregard gap risk, we can consider the following approximations to remove losses, as described in \cite{BrigoPallavicini2013}.
\[
C^k_t \doteq V^k_t\,.
\]%

We can apply this approximation to most quoted interest-rate products, as shown in \cite{BrigoCapponiPallaviciniPapatheodorou}. Furthermore, we could extend it also to FX products, provided that we can discard the devaluation of the FX rate on default events. See for details \cite{Schoenbucher2006}.

Under that assumption we obtain from \eqref{eq:securitypricing} by means of the Feynman-Kac theorem that
\begin{equation}
\label{eq:perfect}
V^k_t = \int_t^T \Ex{t}{ D(t,u;r) \left( d\pi^k_u + V^k_u (r_u - c^k_u) \,du \right) } = \int_t^T \Ex{t}{ D(t,u;c^k) \,d\pi^k_u }
\end{equation}%
where we assume the terminal condition $V^k_T\doteq 0$, and the expectation is taken under the $\mathbb Q$ measure, and we define the discount factor for a generic rate $x_t$ as
\[
D(t,T;x) := \exp\left\{-\int_t^T du\, x_u\right\}\,.
\]%

We notice that the perfect collateralization condition implies that the security prices do not depend any longer on the risk-free rate $r_t$. See \cite{Perini2011} for a discussion of such invariance property.

\subsection{Dividend Processes in Foreign Currencies}

We can extend the previous analysis by assuming that the cumulated dividend processes may be paid or received in currencies different from the currency of the Treasury bank account. We name domestic the currency of the bank account, and foreign all the other currencies.

We introduce for each security $V^k_t$ a set of cumulative dividend processes $D^{k,a}_t$ paying or receiving cash in currency $a$. We define them as
\[
D^{k,a}_t := \pi^{k,a}_t + C^{k,a}_t - \int_0^t c^{k,a}_u C^{k,a}_u \,du
\]%
while the rescaled collateral and cumulative dividend process are given by
\[
{\bar C}^k_t := \sum_a \frac{\chi^a_t}{B_t} \,C^{k,a}_t
\;,\quad
{\bar D}^k_t := \sum_a D^{k,a}_0 + \sum_a \int_0^t \frac{\chi^a_u}{B_u} \,dD^{k,a}_u
\]%
where $\chi^a_t$ is the FX spot rate which converts cash expressed in currency $a$ into the bank account currency. The rescaled gain processes is defined as before as
\[
{\bar G}^k_t := {\bar V}^k_t + {\bar D}^k_t - {\bar C}^k_t\,.
\]

Then, if we assume that the rescaled gain processes are martingales under the $\mathbb Q$ measure, we can compute the price of a security. We follow the algebra in details.
\begin{eqnarray*}
V^k_t &=& B_t \,\Ex{t}{ \frac{V^k_T}{B_T} + \sum_a \int_t^T \frac{dD^{k,a}_u}{B_u} - \sum_a \frac{\chi^a_T C^{k,a}_T}{B^T} + \sum_a \frac{\chi^a_t C^{k,a}_t}{B_t} } \\
      &=& B_t \,\Ex{t}{ \frac{V^k_T}{B_T} + \sum_a \int_t^T \frac{\chi^a_u}{B_u} \,d\pi^{k,a}_u + \sum_a \int_t^T \frac{\chi^a_u}{B_u} \,dC^k_u - \sum_a \frac{\chi^a_T C^{k,a}_T}{B^T} + \sum_a \frac{\chi^a_t C^{k,a}_t}{B_t} + \sum_a \int_t^T \frac{c^k_u \chi^a_u C^{k,a}_u}{B_u} \,du }\\
      &=& B_t \,\Ex{t}{ \frac{V^k_T}{B_T} + \sum_a \int_t^T \frac{\chi^a_u}{B_u} \,d\pi^{k,a}_u + \sum_a \int_t^T \chi^a_u \frac{C^{k,a}_u}{B_u} \left( (c^k_u - r_u) \,du + \frac{d\chi^a_u}{\chi^a_u} \right) }\,.
\end{eqnarray*}%
We define the collateralized basis-curve short-rate $b^a_t(c)$ for currency $a$ and domestic collateralization $c_t$ as
\[
b^a_t(c) \,dt := c_t \,dt - \Ex{t}{\frac{d\chi^a_t}{\chi^a_t}}
\]%
to obtain the pricing equation for a security with foreign dividends
\begin{equation}
\label{eq:foreignsecuritypricing}
V^k_t = B_t \,\Ex{t}{ \frac{V^k_T}{B_T} + \sum_a \int_t^T \frac{\chi^a_u}{B_u} \,d\pi^{k,a}_u + \sum_a \int_t^T \chi^a_u \frac{C^{k,a}_u}{B_u} (c^k_u - b^a_u(c) + c_u - r_u) \,du }\,.
\end{equation}%

A relevant case is given by a derivative contract perfectly collateralized with a margining procedure in a single foreign currency $f$. In such case we can set
\[
\chi^f_t C^{k,f}_t \doteq V^k_t\,.
\]%
If we plug this relationship into equation \eqref{eq:foreignsecuritypricing}, and we apply the Feynman-Kac theorem, we get
\begin{equation}
\label{eq:perfectforeign}
V^k_t = \int_t^T \Ex{t}{ D(t,u;c^k-b^f(c)+c) \sum_a \chi^a_u \,d\pi^{k,a}_u }
\end{equation}%
where we assume the terminal condition $V^k_T\doteq 0$.
%
%
\section{Estimating MtM leg corrections}
In this section we work out the formulae obtained in Section \ref{sec:MtMLegs} and needed to value the NPVs of marked-to-market CCS legs. Our aim is to give a rough yet simple estimate of those contributions and we therefore base our calculations on straghtforward market models with a very few input data to be used.
\subsection{Domestic MtM}\label{sec:DomM2MEstimate}
To value terms entering the NPV of a domestic leg with renotioning, see Eq.\eqref{eq:MtMCCS_2}, we propose a simple market model for forward exchange rates and forward Libor rates with static discounting-forwarding spread.\\ 
Let $E_t(T_i;e)$ be the simple-compounding discounting rate defined as
$$
1+\tau E_t(T_i;e) := \frac{P_t(T_{i-1};e)}{P_t(T_{i};e)}\,,
$$
whose variance drives the Radon-Nikodym derivative needed to link measures $\mathbb{Q}^{T_{i-1};e}$ and $\mathbb{Q}^{T_{i};e}$, see Eq.\eqref{eq:dQidQim1}. In order to cope with negative Libor rates, we assume that forward Libor rates $F_t(T_i;e)$ evolve according to shifted lognormal diffusion processes 
with displacements $\Delta_i\geq 0$
\begin{equation}\label{eq:MMSDEforFd}
dF_t(T_i;e)=\left(F_t(T_i;e)+\Delta_i\right)\eta_idW^{F,i}_t+(\cdots)dt
\end{equation}
where the drift term for $i$-th forward vanishes under $\mathbb{Q}^{T_{i};e}.$
We moreover assume static additive spread between $E_t(T_i;e)$ and $F_t(T_i;e)$ as 
$$
F_t(T_i;e)\approx E_t(T_i;e)+(F_0(T_i;e)-E_0(T_i;e)):=E_t(T_i;e)+\beta_i
$$
such that, under  $\mathbb{Q}^{T_{i};e}$,
\begin{equation}\label{eq:MMSDEforEd}
dE_t(T_i;e)\approx\left(E_t(T_i;e)+\Delta_i+\beta_i\right)\eta_idW^{F,i}_t.
\end{equation}
This choice is quite natural as markets do not quote any volatility product linked to simple-compounding discounting rates.
Now we model forward exchange rates $X_t(T_{i-1};e)$ by lognormal processes getting, under  $\mathbb{Q}^{T_{i};e},$  
\begin{equation}\label{eq:MMSDEforX_1}
dX_t(T_{i-1};e)=X_t(T_{i-1};e)\sigma_{i-1}\left[dW^{X,i-1}_t-\frac{\tau_i\left(E_t(T_i;e)+\Delta_i+\beta_i\right)}{1+\tau_iE_t(T_i;e)}\eta_i\rho^{F,X}_{i,i-1}dt\right]\,,
\end{equation}
where $\rho^{F,X}_{i,i-1}$is the instantaneous correlation between the Brownian motions that drive the $i$-th domestic forward Libor and the $(i-1)$-th forward exchange rate respectively. As we only want to provide simple formulae, we freeze the drift of $X_t(T_{i-1};e)$ at time $t$ and get
\begin{eqnarray}\label{eq:convAdjDomMtm}
\ExT{t}{T_i;e}{\chi_{T_{i-1}}} &=& X_t(T_{i-1};e)\exp\left\{-\frac{\tau_i\left(E_t(T_i;e)+\Delta_i+\beta_i\right)}{1+\tau_iE_t(T_i;e)}\sigma_{i-1}\eta_i\rho^{F,X}_{i,i-1}\left(T_{i-1}-t\right)\right\} \nonumber\\
\ExT{t}{T_i;e}{ \chi_{T_{i-1}}L^d_{T_{i-1}}(T_i)} &=&\ExT{t}{T_i;e}{\chi_{T_{i-1}}}\left[\left(F_t(T_i;e)+\Delta_i\right)\mathrm{e}^{\sigma_{i-1}\eta_i\rho^{F,X}_{i,i-1}(T_{i-1}-t)}-\Delta_i\right]\,.
\end{eqnarray}
\subsection{Foreign MtM}\label{sec:forM2MEstimate}
In utter analogy with the domestic MtM case, we introduce $E^f_t(T_i;e)$, the simple-compounding basis foreign discounting rate defined as
$$
1+\tau E^f_t(T_i;e) := \frac{P^f_t(T_{i-1};e)}{P^f_t(T_i;e)}\,,
$$
which is by construction a $\mathbb{Q}^{T_{i};b}$ martingale, and whose product by $X_t(T_{i-1};e)$ drives the Radon-Nikodym derivative $Z_t^{i-1,i}(e;b)$, see Eq.\eqref{eq:dQibasisdQim1}. We allow the presence of negative foreign Libor rates assuming that basis forward Libor rates $F^f_t(T_i;e)$ evolve according to shifted lognormal diffusion processes 
with displacements $\Delta^f_i\geq 0$
\begin{equation}\label{eq:MMSDEforFf}
dF^f_t(T_i;e)=\left(F^f_t(T_i;e)+\Delta^f_i\right)\eta^f_idW^{F^f,i}_t+(\cdots)dt
\end{equation}
where the drift term for $i$-th forward vanishes under $\mathbb{Q}^{T_{i};b}.$
We moreover assume static additive spread between $E^f_t(T_i;e)$ and $F^f_t(T_i;e)$ as 
$$
F^f_t(T_i;e)\approx E^f_t(T_i;e)+(F^f_0(T_i;e)-E^f_0(T_i;e)):=E^f_t(T_i;e)+\beta^f_i
$$
such that, under  $\mathbb{Q}^{T_{i};b}$,
\begin{equation}\label{eq:MMSDEforEf}
dE^f_t(T_i;e)\approx\left(E^f_t(T_i;e)+\Delta^f_i+\beta^f_i\right)\eta^f_idW^{F^f,i}_t.
\end{equation}
Now we model forward exchange rates $X_t(T_{i-1};e)$ by driftless lognormal processes under  $\mathbb{Q}^{T_{i-1};e},$ such that, switching to measure $\mathbb{Q}^{T_{i};b},$ we have  
\begin{equation}\label{eq:MMSDEforX_2}
dX_t(T_{i-1};e)=X_t(T_{i-1};e)\sigma_{i-1}\left[dW^{X,i-1}_t+\left(\sigma_{i-1}-\frac{\tau_i\left(E^f_t(T_i;e)+\Delta^f_i+\beta^f_i\right)}{1+\tau_iE^f_t(T_i;e)}\eta^f_i\rho^{F^f,X}_{i,i-1}\right)dt\right]\,,
\end{equation}
and, by It\^o's formula,
\begin{equation*}
\frac{d\left(1/X_t(T_{i-1};e)\right)}{1/X_t(T_{i-1};e)}=\sigma_{i-1}\left[\frac{\tau_i\left(E^f_t(T_i;e)+\Delta^f_i+\beta^f_i\right)}{1+\tau_iE^f_t(T_i;e)}\eta^f_i\rho^{F^f,X}_{i,i-1}dt-dW^{X,i-1}_t\right]\,,
\end{equation*}
where $\rho^{F^f,X}_{i,i-1}$is the instantaneous correlation between Brownian motions driving the $i$-th foreign basis forward Libor and the $(i-1)$-th forward exchange rate. By freezing the drift of $X_t(T_{i-1};e)$ at time $t$ we get
\begin{eqnarray}\label{eq:convAdjForMtm}
\ExT{t}{T_i;b}{\frac{1}{\chi_{T_{i-1}}}} &=& \frac{1}{X_t(T_{i-1};e)}\exp\left\{\frac{\tau_i\left(E^f_t(T_i;e)+\Delta^f_i+\beta^f_i\right)}{1+\tau_iE^f_t(T_i;e)}\sigma_{i-1}\eta^f_i\rho^{F^f,X}_{i,i-1}\left(T_{i-1}-t\right)\right\}\nonumber\\
\ExT{t}{T_i;b}{\frac{L^f_{T_{i-1}}(T_i)}{\chi_{T_{i-1}}}} &=&\ExT{t}{T_i;b}{\frac{1}{\chi_{T_{i-1}}}}\left[\left(F^f_t(T_i;e)+\Delta^f_i\right)\mathrm{e}^{-\sigma_{i-1}\eta^f_i\rho^{F^F,X}_{i,i-1}(T_{i-1}-t)}-\Delta^f_i\right]\,.
\end{eqnarray}
\begin{remark}{\bf (Alternative approach)}
These formulae are only formally analogous to those of the domestic MtM case, but the dynamical choices we made are different. We could however have decided to exploit the same modelling assumptions as in Sec.\ref{sec:DomM2MEstimate}, by switching back to domestic measures in Eq.\eqref{eq:VCCSf_mtm}, getting 
\begin{eqnarray}
\ExT{t}{T_i;b}{\frac{1}{\chi_{T_{i-1}}}} &=& \ExT{t}{T_i;e}{\frac{X_{T_i}(T_i;e)}{X_{T_{i-1}}(T_{i-1};e)}}\nonumber\\
\ExT{t}{T_i;b}{\frac{L^f_{T_{i-1}}(T_i)}{\chi_{T_{i-1}}}} &=& \ExT{t}{T_i;e}{\frac{X_{T_i}(T_i;e)F^f_{T_{i-1}}(T_i;e)}{X_{T_{i-1}}(T_{i-1};e)}}\,.
\end{eqnarray}
The first contribution could be easily obtained by resorting to the same model as in Eqs.(\ref{eq:MMSDEforEd}-\ref{eq:MMSDEforX_1}) and it would involve the correlation between consecutive forward FX rates and the correlation  between the $(i-1)$-th forward exchange rate and the $i$-th domestic discount simple rate. The second term would require to define the dynamics of $F^f_t(T_i;e)$ under $\mathbb{Q}^{T_i,e}.$ The result would be a little bit more involved than the one of Eq.\eqref{eq:convAdjForMtm}, but still easy to obtain, as the Radon-Nikodym derivative linking $\mathbb{Q}^{T_i,b}$ to $\mathbb{Q}^{T_i,e} $ reads
$$
\left.\frac{d\mathbb{Q}^{T_i,b}}{d\mathbb{Q}^{T_i,e}}\right|_t=\frac{X_t(T_i;e)}{X_0(T_i;e)}\,.
$$
It is worth stressing that, because of the identity 
$$
\frac{X_{t}(T_i;e)}{X_{t}(T_{i-1};e)}\equiv\frac{1+\tau_iE_t(T_i;e)}{1+\tau_iE^f_t(T_i;e)}\,,
$$
the dynamics of $E_t(\cdot;e),E^f_t(\cdot;e),X_t(\cdot;e)$ are intimately entangled. If we set as fundamental drivers the forward exchange rates and the domestic simple rates, then the dynamics of the foreign basis simple rates will follow by It\^o's rule, and will no longer correspond to a free choice. For instance, according to Eqs.(\ref{eq:MMSDEforEd}-\ref{eq:MMSDEforX_1}), we get  
\begin{multline*}
dE^f_t(T_i;e)=\left(\frac{1}{\tau_i} + E^f_t(T_i;e)\right) \left(\sigma_{i-1}dW^{X,i-1}-\sigma_{i}dW^{X,i}+\frac{\tau_i\left(E_t(T_i;e)+\Delta_i+\beta_i\right)}{1+\tau_iE_t(T_i;e)}\eta_idW^{F,i}\right)\,.
\end{multline*}
In such a framework it would be unnatural to think that the $F^f_t(\cdot;e)$ and the $E^f_t(\cdot;e)$ are related by a static spread. 
\end{remark} 
\begin{remark}{\bf (Foreign basis Libor rates volatility)}
As  there are no market products from which we can directly derive foreign basis forward Libor rate  volatilities $\eta^f_i$, we suggest, as a reasonable proxy, to use ATM caplet/floorlet implied volatilities deduced from single-currency foreign on-shore market.     
\end{remark}
\newpage

\bibliographystyle{plainnat}
\bibliography{fx_collateral}

\end{document}